# The Z1: Architecture and Algorithms of Konrad Zuse's First Computer

Raul Rojas
Freie Universität Berlin

Abstract

This paper provides the first comprehensive description of the Z1, the mechanical computer built by the German inventor Konrad Zuse in Berlin from 1936 to 1938. The paper describes the main structural elements of the machine, the high-level architecture, and the dataflow between components. The computer could execute the four basic arithmetic operations using binary floating-point numbers. Instructions were read from punched tape. A program consisted of a sequence of arithmetical operations, intermixed with memory store and load instructions, interrupted possibly by input and output operations. Numbers were stored in a mechanical memory. The machine did not include conditional branching in the instruction set.

While the architecture of the Z1 is similar to the relay computer Zuse finished in 1941 (the Z3) there are some significant differences. The Z1 implements operations as sequences of microinstructions, as in the Z3, but does not use rotary switches as micro-steppers. The Z1 uses a digital counter and a set of conditions which trigger microinstructions for the exponent and mantissa units, as well as for the memory blocks. Microinstructions select one out of 12 layers in a machine with a 3D mechanical structure of binary mechanical elements. The exception circuits for mantissa zero, necessary for normalized floating-point, were lacking; they were first implemented in the Z3.

The information for this article was extracted from careful study of the blueprints drawn by Zuse for the reconstruction of the Z1 for the German Technology Museum in Berlin, from some letters, and from sketches in notebooks. Although the machine has been in exhibition since 1989 (non-operational), no detailed high-level description of the machine's architecture had been available. This paper fills that gap.

## 1 Konrad Zuse and the Z1

The German inventor Konrad Zuse (1910-1995) built his first computing machine from 1936 to 1938[3] (from 1934 to 1935 he experimented with small mechanical circuits). In Germany, Zuse has always been considered the father of the computer although the machines he built during WWII became known only after the conflagration. Zuse studied civil engineering at the Technische Hochschule Charlottenburg (today's Technical University of Berlin). His first employer was the company *Henschel Flugzeugwerke*, who had just started building military airplanes in Berlin in 1933 [1]. The duty of the 25 years old was to carry out the long chains of structural calculations needed for the manufacturing process of aircraft components. As a student, Zuse had already started thinking about ways of mechanizing computations [2]. Therefore, after working just several months for the *Henschel Flugzeugwerke*, he decided to quit, build a mechanical computer, and start his own business, in fact, the first computer company in the world.

---

[1] Version 2 of this paper modifies slightly the Z1 datapath and includes the full instruction table for the operations of the Z1. The microinstruction plates were redrawn to show that only one teeth presses the pins A, B, C or D.

[2] Version 3 further specifies better the datapath, while the instruction set is explained in a more complete way. A full description of the 12 layers of the processor I now included in section 11.

[3] The precise chronology of his line of computing machines was provided by K. Zuse in a small handwritten note from March 1946. There, the V1 is dated as having been built in the years 1936-1938. In another memo from April 1986, he states that the Z1 was finished at the end of 1937 (zuse_archive_203_5 at zuse.zib.de).

During the period 1936-1945, Konrad Zuse was unstoppable, even after two short-lived calls to the front. He could manage to be recalled to Berlin to work part-time for *Henschel,* and part-time for his own company. In those nine years, he built the six computers known today as the Z1, Z2, Z3 and Z4, as well as the specialized S1 and S2 machines. The last four were built after WWII had already started. The Z4 was finished during the closing months of the war. Zuse's original abbreviations for the machines' names were V1, V2, V3 and V4 (meaning "Versuchsmodell", or prototype). After the war, he changed the V for a Z for obvious reasons. The V1 (Z1 in what follows) was a fascinating piece of technical brinkmanship: it was a completely mechanical computer, but instead of using gears and wheels to represent the ten decimal digits (as Babbage had done in the previous century, or IBM was doing with its Hollerith machines), Zuse decided to build a fully binary computer. His machine was based on components in which the forward linear movement of a small rod or metallic plate represented a one, and no movement represented a zero (or vice versa, according to the component). Zuse developed novel types of mechanical logical gates and finished the first prototype of the machine in his parent's living room. The sequence of events that led to the Z1 and subsequent machines has been appraised by Zuse himself in his autobiography [2].

The Z1 was a mechanical but also a surprisingly modern computing machine: it was based on binary numbers, it used a floating-point representation for the data and could perform the four basic arithmetic operations. The program was read from a punched tape (no conditional branch was available though), and the results could be stored to or read from memory (16 words). The machine cycle was around 4 Hz.

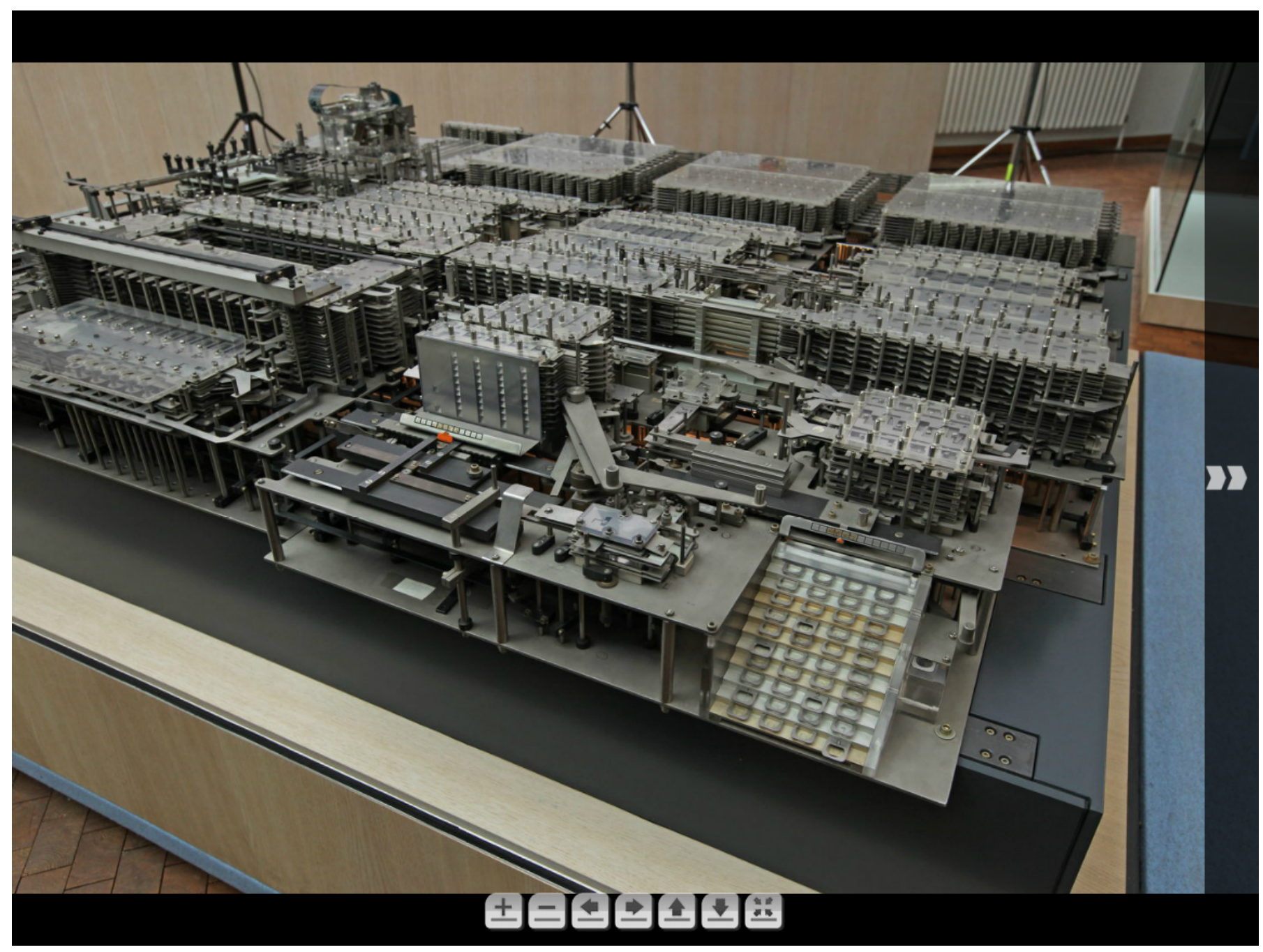

Fig. 1: A view of the reconstructed Z1 in Berlin (from the Konrad Zuse Internet Archive [5]). The user can rotate the view around the machine, can zoom in and out. The virtual display is based on thousands of linked photographs.

The Z1 was very similar to the Z3, finished in 1941, whose architecture has been described in detail in the IEEE Annals [3]. However, the detailed high-level architecture of the Z1 has never been explained before. The original prototype was destroyed during a bombing raid in 1943. Only a few sketches and some photographs of the mechanical components survived. In the 1980s, Konrad Zuse, who had retired many years earlier, obtained financing from Siemens and other German sponsors for building a full replica of the Z1 which is now housed in Berlin's Technology Museum (Fig. 1). Zuse built the machine with the help of two engineering students: He prepared a full set of blueprints, painstakingly drawing every single mechanical component (to be cut from sheets of steel), and supervising the construction over the course of several years at his own house in Hünfeld, Germany. The first sketches of the Z1 reconstruction were drawn in 1984. In April of 1986 Zuse drew a timeline expecting to have the machine finished by December of 1987. When the machine was delivered to the Berlin museum in 1989 it was shown running and computing some arithmetical operations on several occasions. However, the reconstructed Z1 was, like the original, never reliable enough to run unattended for long stretches of time. It even failed at the inauguration and Zuse spent months repairing the machine. After Konrad Zuse passed away in 1995, the machine was never restarted again.

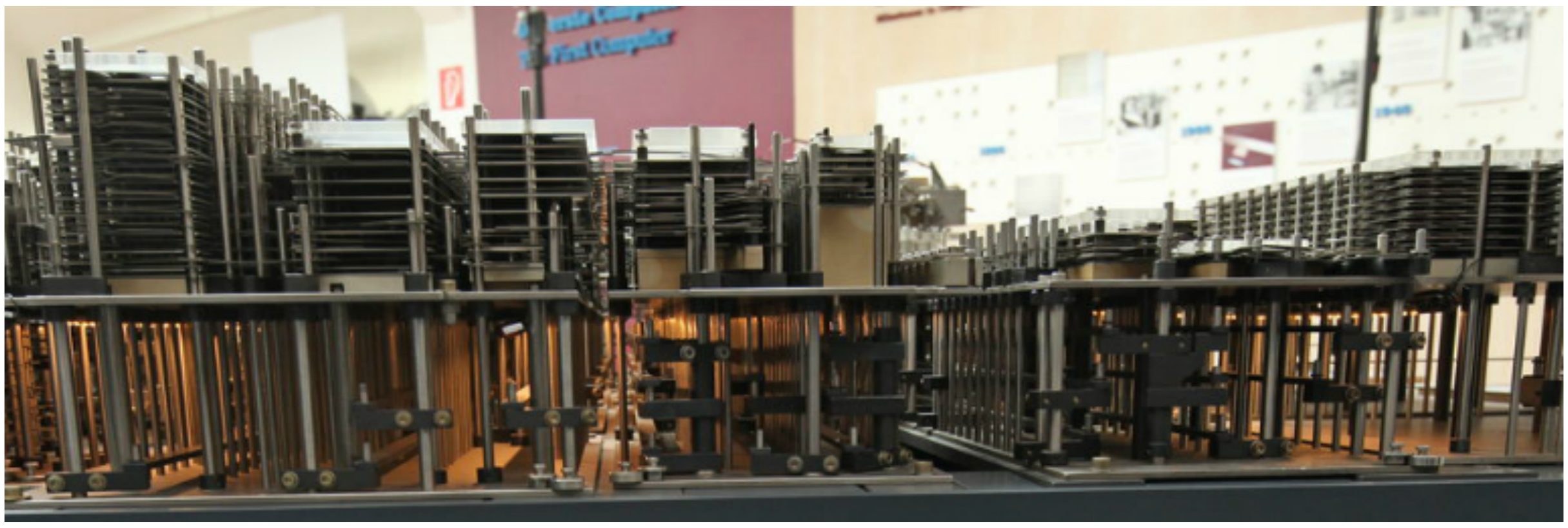

Fig. 2: The mechanical layers of the Z1. The eight memory layers can be seen on the right, the 12 processor layers on the left. The lower section with levers is used to coordinate all parts of the machine. Zuse called it "the basement".

Although we have a reconstruction of the Z1 in Berlin, fate struck twice. Other than drawing the blueprints, Zuse made no serious effort to write a complete top-down description of the reconstructed Z1 (he expected a local university to do it). This would have been necessary, because it is evident from comparing the reconstruction with old photographs of the Z1 built in 1938, that the new machine has been "streamlined". The higher precision of the machining equipment available to Zuse in the 1980s allowed him to build the reconstruction using layers of steel plates which could be placed tighter together. The new Z1 fills a significantly smaller volume than the old Z1. It is also not completely clear if the new Z1 is strictly a one-to-one logical and mechanical clone of the original machine, or if Zuse's experience with the Z3 and later machines allowed him to improve portions of the reconstructed Z1. In the set of mechanical blueprints drawn from 1984 to 1989, there are at least *six different drafts* for the addition unit, having between five to eight, and finally up to 12 mechanical layers.[4] Zuse left no detailed written record which could allow us to answer such questions. Still worse, he rebuilt the Z1 and left no comprehensive logical description of it – for the second time! He acted like those celebrated clockmakers who only draw the parts of their watches, leaving no further explanation: first-rate clockmakers would need no further clarifications. His two student assistants documented only the memory and the tape reader, an extremely useful piece of information [4]. Visitors to the Berlin Technology Museum can only wonder at the thousands of components visible in the machine. They can both wonder and despair, since it is almost impossible, even for professional computer scientists, to visualize the inner workings of this mechanical Leviathan. The machine has been there since 1989 -- but unfortunately dead.

[4] All the blueprints for the reconstruction of the Z1 have been made available through our „Konrad Zuse Internet Archive" at http://zuse-z1.zib.de.

This paper is based on a careful study of the blueprints of the Z1, scattered notes in Zuse's notebooks, and numerous on-site inspections of the machine. The reconstructed Z1 has been non-operational for so many years because the steel plates used by Zuse bend under stress. For this paper, more than 1100 large format drawings of the machine's components were reviewed, as well as 15.000 pages in notebooks (only a small fraction thereof contained information about the Z1 though). I could only see a short video of parts of the machine operating (filmed almost 20 years ago). Deutsches Museum in Munich houses 1079 blueprints from Zuse's private papers, while the Berlin Technical Museum has another 314 in its archives. Fortunately, some blueprints include also the definition and timing of some microinstructions for the Z1, and also a few examples of bit-by-bit handwritten calculations made by Zuse. Such examples were probably used by Zuse to check the machine's internal operation and find bugs. This information was like a Rosetta stone, which allowed us to correlate the Z1 microinstructions with the diagrams and blueprints, and also with our relatively deep knowledge of the relay-computer Z3 (for which we have complete circuits [5]). The Z3 is based on the same high-level architecture as the Z1, but is different in an important number of ways.

This paper proceeds top-down: first we review the block architecture of the Z1, the layout of the mechanical components, and I also provide some examples of the mechanical gates used by Zuse. We then look in more detail at the Z1 core elements: the clocked addition units for exponent and mantissa, the memory, and the microsequencer for arithmetical operations. We show the interplay of the mechanical elements and how the "sandwiched" layout of steel plates helped Zuse organize the computation. We look at addition and subtraction, at the multiplication and division process, and also at input and output. The last part of the paper briefly situates the Z1 in its historical context.

## 2 Binary coding and floating point

The Z1 uses binary coding for all numbers stored in memory and handled by the processor. Negative numbers are represented using two's complement arithmetic. The Z1 uses binary floating point with normalized mantissas. That is: a number $n$ is represented using $k$ bits $b_k, b_{k-1}, \dots, b_0$ in such a way that

$$n = b_k 2^k + b_{k-1} 2^{k-1} + \cdots + b_0 2^0$$

A negative number $-n$ is represented by

$$-n = \overline{b_k} 2^k + \overline{b_{k-1}} 2^{k-1} + \cdots + \overline{b_0} 2^0 + 1$$

Where the bits $\overline{b_\ell}$ are the logical complement of the bits $b_\ell$ in the binary representation of $n$. Notice that we can produce the negative of a number $n$ by complementing its bits, and adding 1 at the end to the whole binary representation. This is, in fact, what the Z1 does when negating a number, delaying the addition of 1 until the complement of $n$ passes through the addition unit, where the 1 is added.

The Z1 uses 7 bits for the exponent in two's complement arithmetic, giving it a representation range from $-64$ to $+63$.

For the floating-point (FP) representation, the Z1 handles a number as $2^a \cdot b$, where $a$ is the binary exponent and $b$ the binary mantissa. The mantissas are "normalized", that is, they have a 1 before the binary point. For example, $b = 1.11$ represents the normalized mantissa $1 + 1/2 + 1/4$. In general, a normalized mantissa has the form:

$$b = b_0 2^0 + b_{-1} 2^{-1} + \cdots + b_{-k} 2^{-k}$$

Where $b_0 = 1$, and the other bits can be zero or one. Note that with this convention, $b = 0$ cannot be represented. The Z3 used the convention that the lowest exponent ($a = -64$ in this case) represented the number zero, regardless of the mantissa bits. This was not implemented in the reconstructed Z1, and we discuss later what problems this provoques.

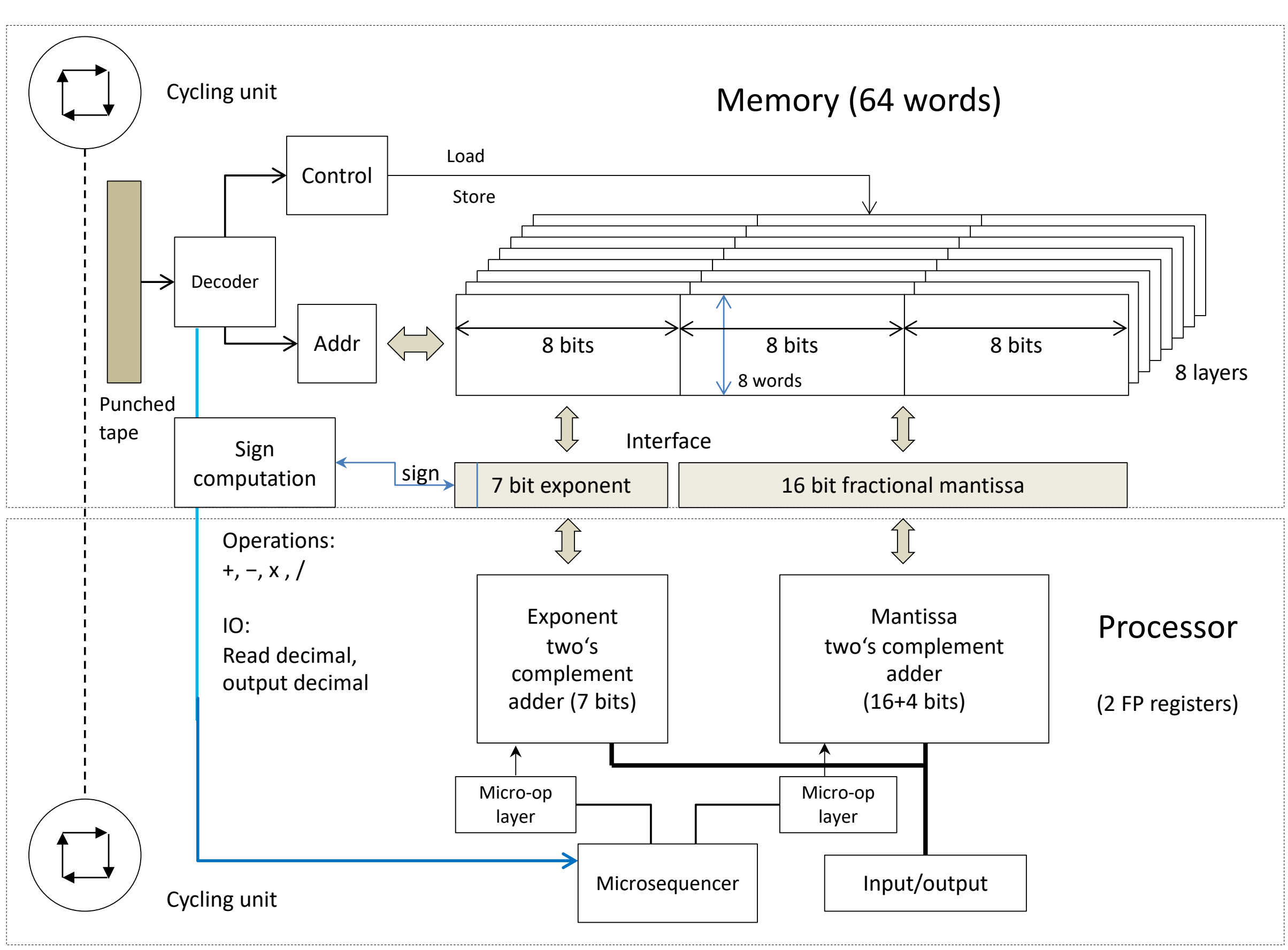

Fig. 3: Block diagram of the Z1 (1936-38) according to the reconstruction of 1989. The original Z1 had only 16 words of memory instead of 64. The punched tape was made of 35mm film tape. Each instruction was encoded using 8 bits.

A FP number with a non-normalized mantissa can be normalized by shifting the mantissa and adjusting the binary exponent. For example, $2^1 \times 0.11 = 2^0 \times 1.1$. The first representation is not normalized, the second is, although in both cases we have the same number. This normalization is used in the Z1 after some arithmetical operations in order to keep normalized mantissas. One advantage of normalized mantissas is that the leading bit in the mantissa does not need to be stored in memory, since it is always one. That saves on bit for each word, and when you only have sixteen of those, that is important.

## 3 Block-architecture

The Z1 was a clocked machine. Being a mechanical device, the clock was subdivided into four subcycles which consisted in the movement of mechanical components in four orthogonal directions, as shown in Fig. 2 (left side, see "cycling unit"). Each movement direction was called an "engagement" by Zuse. He aimed for a 4Hz clock cycle but the Berlin reconstruction never was operated faster than at 1Hz (four engagements per second). At that speed, a multiplication takes around 20 seconds.

The diagram in Fig. 3 is anachronistic, in the sense that it was never drawn like this by Zuse, but in terms of modern computer architecture, this diagram shows the main components and their interplay. The Z1 has a number of features later adopted in the Z3. From a modern perspective, the most important innovations in the Z1 (see Fig. 3) were the following:

- It was based on a fully binary architecture for the memory and the processor.
- The memory was separated from the CPU. In the Berlin reconstruction, the memory and punched tape reader constitute about one half of the machine. The processor, I/O consoles, and the microcontrol unit constitute the other half. The original Z1 had 16 words of memory, the reconstruction has 64.
- The machine was programmable: the program was read from punched tape using 8 bits (two bits for the opcode and six bits for memory addressing, or three bits for the opcode of the four arithmetical and the two I/O operations). Therefore, there were only eight instructions: the four basic arithmetical operations ($+,-,\times,\div$), load-from and store-to memory, one instruction ("up") for reading data from a decimal panel for decimal-binary conversion, and another for showing the contents of the result register on a mechanical decimal display ("down").
- Floating-point was used for internal data representation, in the memory and in the processor. Therefore, the processor was divided into two parts: one for handling the exponents, another for handling the mantissas. The mantissa had 16 bits for the bits after the binary point (in memory). Two additional bits were used in the processor, to increase the accuracy to 18 bits after the point. The bit to the left of the point was always 1 (normalized floating-point) and did not have to be stored. In the processor this bit and one more ($b_0, b_1$) were used to the left of the binary point. Exponents were represented with 7 bits in two's complement format (running thus from $-64$ to $+63$). The sign of the floating-point numbers was stored in one additional bit. Therefore, the word-length in memory was 24 bits (16 bits for the mantissa, 7 for the exponent, one bit for the sign). In the processor 20 bits were used to handle mantissas.
- As said before, the special case of zero in arguments or results (which cannot be expressed with a normalized mantissa, where the leading bit is always 1) can be handled within the floating-point representation as special values of the exponent. This was done in the Z3 *but not in the Z1*, *also not in its reconstruction*. Therefore, neither the original Z1, nor the reconstruction, can work with zero as initial or final result (although the machine can continue working with zero results during some of the algorithms). Zuse was aware of this shortcoming, but he left the solution to the relay machine of 1941, which was easier to wire.

- The CPU was microcoded: operations were broken into sequences of microinstructions, one for each machine cycle. The microinstructions produced a specific dataflow within the arithmetical-logical units (ALUs), which were running nonstop, adding whatever two numbers were stored in its two input registers, in every cycle.
- Curiously, memory and processor ran independently: the memory would put data at or collect data from the communication interface, whenever the punched tape gave the order. The processor would pick, or put data at the interface, when a load or store operation was executed. It was possible to run only the processor and shut-down the memory, in which case the data at the interface, supposedly coming from the memory, became zero. It was also possible to run only the memory and shut-down the processor. This allowed Zuse to debug each half of the machine independently. When running together, a shaft connecting the cycling units in each half synchronized both parts of the machine.

Further innovations in the Z1 were similar to some of the ideas presented later in the Z3. The instruction set was practically the same as in the Z3 but the Z1 could not extract square roots. The Z1 used discarded 35mm film tapes as punched tape.

Fig. 3 shows the abstract diagram of the reconstructed Z1. Notice the two main halves of the machine: the memory is in the upper half, and the processor in the bottom half. Each half had its own rotating cycling unit, which further divided each cycle into four mechanical movements in the directions shown by the arrows. These four movements could be communicated to any part of the machine using the levers distributed under the computational components. The punched tape was read, one instruction at a time. The instructions had different durations. Load and store operations took one cycle, all other operations needed several cycles. The memory address was contained in the lower six bits of the 8-bit opcode, allowing the programmer to refer explicitly to 64 memory addresses.

Memory and processor communicated through the buffer between both units shown in Fig. 2. In the CPU, the internal representation of the mantissa was extended to 20 bits: two additional bits were added before the binary point (for the binary powers $2^1$ and $2^0$), and two more bits for the lowest binary powers ($2^{-17}$ and $2^{-18}$), in order to increase the accuracy of the CPU for intermediate results. In total, im the processor the mantissa had 20 bits representing the binary powers $2^1$ to $2^{-18}$.

The decoder took an instruction from the punched tape reader, determined the type of operation, and started controlling the memory unit and the processor as needed. A number could be read from memory to the first of two CPU floating-point registers (using a load operation). A further load operation would read a number from memory to the second CPU register. The two registers could be added, subtracted, multiplied, or divided in the processor. Such operations require exponent addition or subtraction (with a two's complement adder), as well as an adder for the mantissas. The sign of the result of an arithmetic operation was handled in a "sign unit" connected directly to the decoder and the microcontrol unit.

An input instruction ("up") from the punched tape stopped the machine and allowed the operator to enter data by pulling four decimal digits from a mechanical panel, entering the exponent of the floating-point representation with a small slider, and also the sign of the number. With the slider, the decimal exponents ran for $-6$ to $+6$. The operator could then restart the machine. An output instruction stopped also the machine and showed the contents of the result register in a decimal mechanical panel, until the operator restarted the machine pressing a lever. The smallest number that could be entered was $1 \times 10^{-6}$. The largest $9999 \times 10^{6}$, although internally the Z1 could compute with smaller or bigger numbers than that.

The microsequencer in Fig. 3 constitutes, together with the exponent and mantissa addition units, the core of the computation capabilities of the Z1. Each arithmetical or I/O operation was divided into “phases”. The microsequencer started counting the phases, starting from zero, and selected the appropriate microoperation in the corresponding layer, out of 12 possible layers of mechanical components in the addition units. The selection of a microinstruction was triggered by the binary opcode of the instruction, the phase (in binary) and two condition flags. When the opcode, condition flags, and phase corresponded to that encoded in a metallic plate, this plate snapped and triggered the desired microinstruction.

Therefore, a minimal program in a punched tape could be, for example: 1) Load number from address 1 (implicitly to the first CPU register), 2) Load number from address 2 (implicitly to the second CPU register), 3) add, 4) show result in decimal. Such simple programs could thus allow the operator to use the Z1 as a simple mechanical calculator for predefined operations. Of course, the sequence of computations could be much longer: long sequences of operations were programmed in a punched tape using the memory as storage for constants and intermediate results (in the latter Z4 computer, one tape used for mathematical computations was two meters long).

The architecture of the Z1 can be summarized using modern terminology as follows: it was a programmable normalized floating-point Von Neumann machine (processor and memory were separate, although the program was stored on tape), with external read-only program, with a memory for sixteen 24-bit words. It was capable of accepting decimal numbers of four digits (and an exponent, as well as a sign) as input, for transforming them into binary. It was capable of performing the four arithmetical operations with the data. The binary floating-point result could be transformed back into decimal scientific notation readable by the user. There was no conditional, nor unconditional branching in the instruction set. Exception handling for zero results was lacking. Each instruction was broken into microinstructions “hardwired” in the machine. A microsequencer orchestrated the execution of the microinstructions. In the single surviving video of the machine operating, it looks to the eye as the moving parts of a heirloom. But this machine was weaving numbers.

## 4 Layout of the mechanical components

The Berlin reconstruction of the Z1 is based on a very clean layout. All mechanical components seem to have been arranged in an optimal way. We mentioned that Zuse designed at least six different versions of the processor. The relative position of the main components was fixed from the beginning and might reflect the original distribution of the mechanical elements in the original Z1. There are two main divisions: a gap separates the memory from the processor (as shown in Fig. 3). In fact, both parts of the machine could be actually pulled apart for debugging purposes since they were mounted on separate tables with rollers. A further horizontal plane subdivides the machine into an upper part containing the computational components (those visible in photographs of the Z1), and a lower part containing all the synchronization levers. This Z1 “underworld” is only visible when the visitor bends over to look underneath the computational skyline. Fig. 4 is a drawing from the blueprints showing the computation and the synchronization levels for part of the processor. Notice the 12 layers of computational components and the lower section with three levels for levers. This blueprint is a good example of how difficult it can be to interpret the drawings. While there is much detail about the size of the parts, there are just a few annotations about their use.

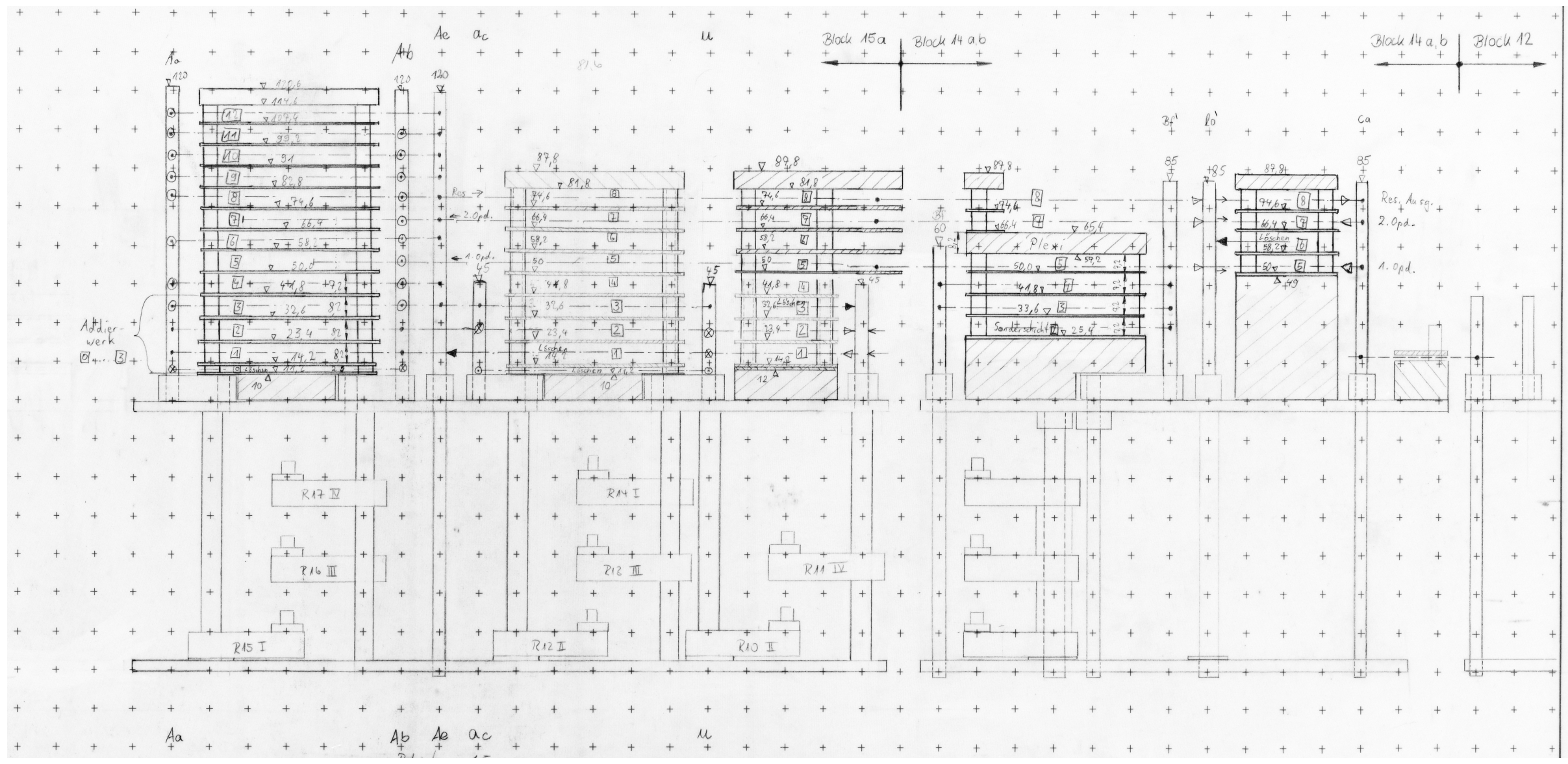

Fig. 4: Blueprint for the computation and synchronization layers of the Z1 (exponent unit)

Fig. 5 shows the distribution of logical components in the reconstructed Z1, seen from the top, and as drawn by Zuse, further annotated with the logical functionality of each block (this sketch has been available since the 1990s). On the upper part we see the three memory banks. Each can contain eight 8-bit words per layer. Every memory bank has 8 mechanical layers, so that a total of 64 words can be stored. The first memory bank (10a) is used for the exponent and sign, the last two banks (10b, 10c) are used for the lower 16 bits of the mantissa of the stored numbers. This distribution of bits allowed Zuse to build three identical 8-bit memory banks and use them for exponent and mantissa, simplifying thus the mechanical construction. Between memory and processor there is a “buffer” for passing numbers to the processor (blocks 12abc), or for receiving numbers from it. There is no way of coding constants in the punched tape. All numbers have to be entered by the user using the decimal input panel (block 18, right side), or must be generated by the computer itself as intermediate results.

Each unit in this diagram shows just the uppermost layer. Remember that the Z1 is built like a “sandwich” of mechanical parts. Each computational layer is strictly separated from the laver above or below (each layer has a metallic floor and a metallic ceiling). Communication between layers is done using vertical rods than can pass movement from one layer to those above or below it. The vertical rods are the small circles drawn between the rectangles representing layers of computation. The somewhat larger circles drawn inside the rectangles represent logical gates. Inside each circle we can find a binary gate (and going down through the layers, up to 12 gates for each circle). This drawing allows us to make an estimation of the number of logical gates present in the Z1. Not all units have the same height, and not all layers are populated with mechanical components. A conservative estimation of the number of binary elements would be 6.000 gates.

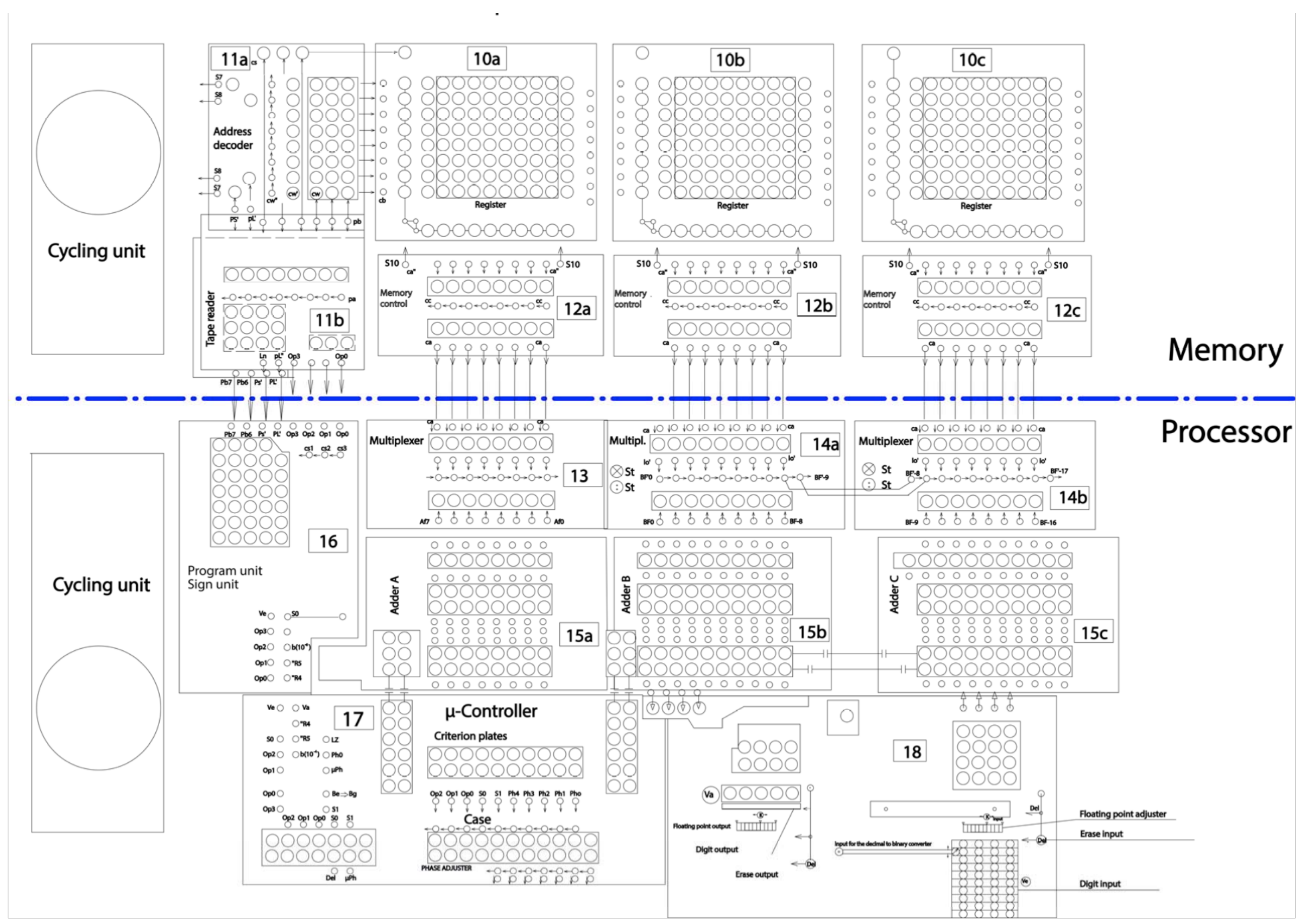

Fig. 5: Diagram of the Z1, showing the mechanical building blocks.

Zuse assigned the numbers shown in Fig. 5 to the different modules of the machine. The purpose of the modules is the following:

Memory Block
11a: Decoder for the six-bit memory addresses
11b: Punched tape reader and op-code decoder
10a: Memory bank for 7-bit exponents and sign
10b, 10c: Memory banks for the fractional part of the mantissa
12abc: Interface to and from the processor for load or store operations

Processor Block
16: Control and sign unit
13: Multiplexer for the two ALU registers in the exponent part
14ab: Multiplexer for ALU mantissa registers, one-bit two-way shifters for multiplication and division
15a: ALU for the exponent
15bc: 20-bit ALU for the normalized mantissa (18 bits for the fractional part)
17: Microcode control
18: Decimal input panel to the right, output panel to the left

One can imagine computation flowing in this diagram from top to bottom: from the memory, the data comes to fill the two registers available to the programmer (which we call F and G). These two registers are distributed along blocks 13 and 14ab. The two registers are fed to the ALUs (blocks 15abc). The result is cycled back to register F or G (as result register), or back to memory. The result can be shown in the decimal display using the "down" instruction (binary to decimal conversion). In what follows we look in more detail to each module, concentrating our efforts in the main computational components.

## 5 The mechanical gates

The mechanical structure of the Z1 can be best understood by looking at a few simple examples of the type of binary logic gates that Zuse used in his machines. The classical digital representation for decimal digits has always been the rotary dial. A gear is divided into ten sectors -- by rotating the gear it is then possible to count from zero to nine. Zuse decided as early as 1934 to use the binary system (which he called, following Leibniz, the dyadic system). In Zuse's technique, a planar plate can have one of two positions (0 or 1). It is possible to move from one state to the other using linear motion. Logical gates pass movement from one plate to another, according to the value of the represented bits. The structures are three-dimensional: they consist of arrangements of superposed planar plates which transmit movement usually through cylindrical rods or pins positioned vertically, at right angles to the plates.

We show examples of the three basic gates: conjunction, disjunction, and negation. There are many possible mechanical realizations for the main idea, and Zuse showed great creativity drawing always the variation of a gate that best fitted the 3D structure of the machine. Fig. 6 shows what Zuse called the "elementary gate". The "actor plate" can be regarded as the machine cycle. This plate moves cyclically from right to left and back. The upper plate is the data bit we are using for control. It can be in the position 1 or 0. The rod going through the openings moves horizontally following the plate (keeping its verticality). If the upper plate is in the 0-position, the actor plate's movement cannot be transmitted to the actuated plate (see Fig. 6, left side). If the data bit plate moves to the 1-position, the movement of the actor plate is transmitted to the actuated plate. This is what Konrad Zuse called a "mechanical relay", just a switch that closes a mechanical "current". This elementary gate can thus copy a bit from the upper to the actuated plate, rotating the movement of the bit by 90 degrees.

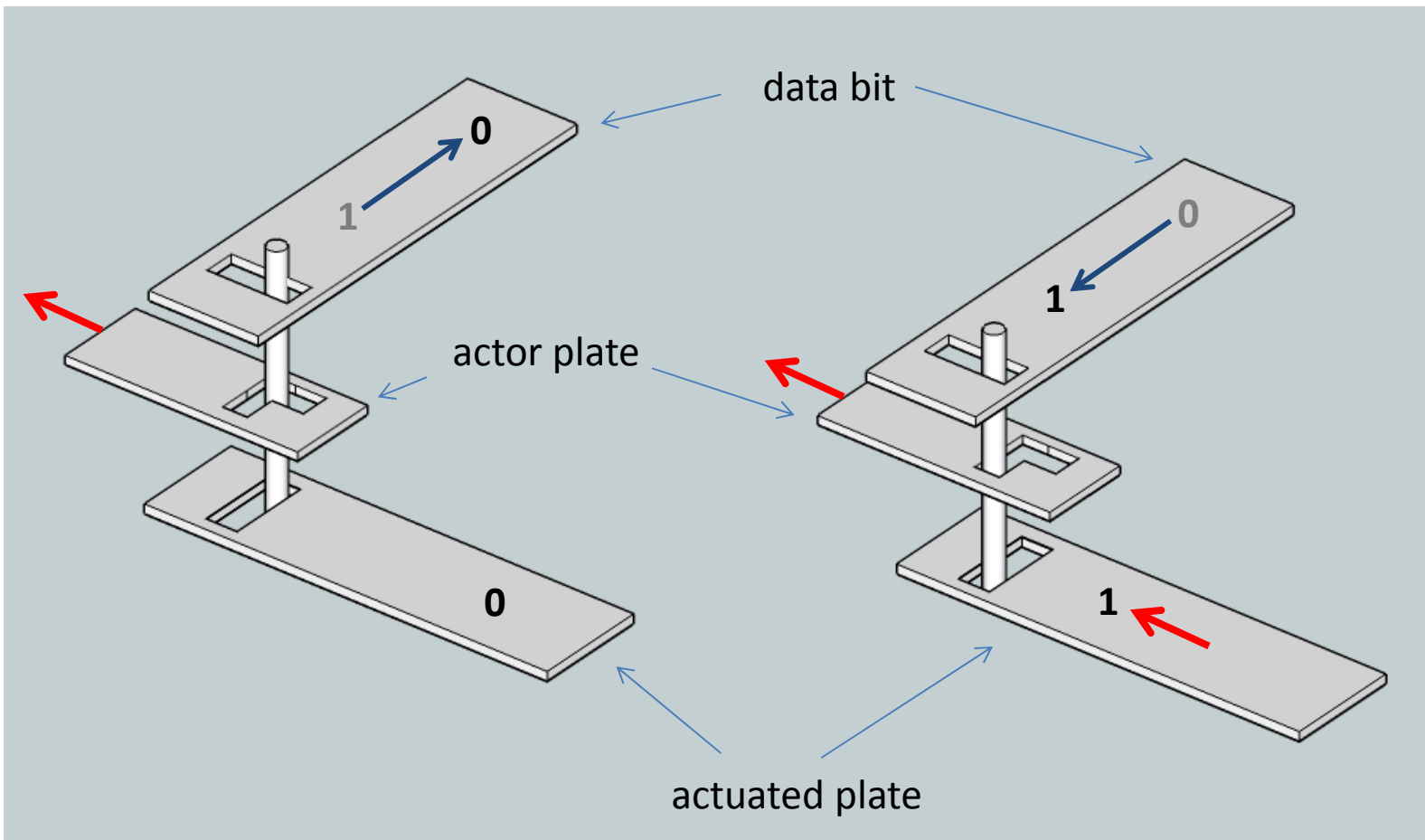

Figure 6: An elementary gate is a switch. If the data bit is 1, the actor and actuated plates are connected. If the data bit is zero, they are disconnected and the movement of the actor is not transmitted.

Fig. 7 shows now such plate arrangements as seen from the top. The actor plate is shown with its opening. The control plate in green pulls the circle (rod) up or down. The actuated plate (red) can move to the right or left, but only when the rod is in such a position that the actor's opening moves the rod. There is a drawing of the equivalent switch to the left of each mechanical top view. The control bit can close or open the gate. The actor plate can be pulled or pushed (as shown by the arrows). Zuse's convention was to always draw the switch in the zero position of the control bit, as done in Fig. 7. Zuse preferred plates to be pushed by the actor plate (right side of Fig. 7) rather than be pulled (left side of Fig. 7). It is now very easy to build a negation gate, by using a closed switch which is opened by the position 1 of the control bit (as shown in the bottom two diagrams in Fig. 7).
Having a mechanical relay, it is now straightforward to build the rest of the logical operations. Fig. 8 shows the necessary circuits, now only using abstract notation. The equivalent mechanical contraptions are easy to conceive.

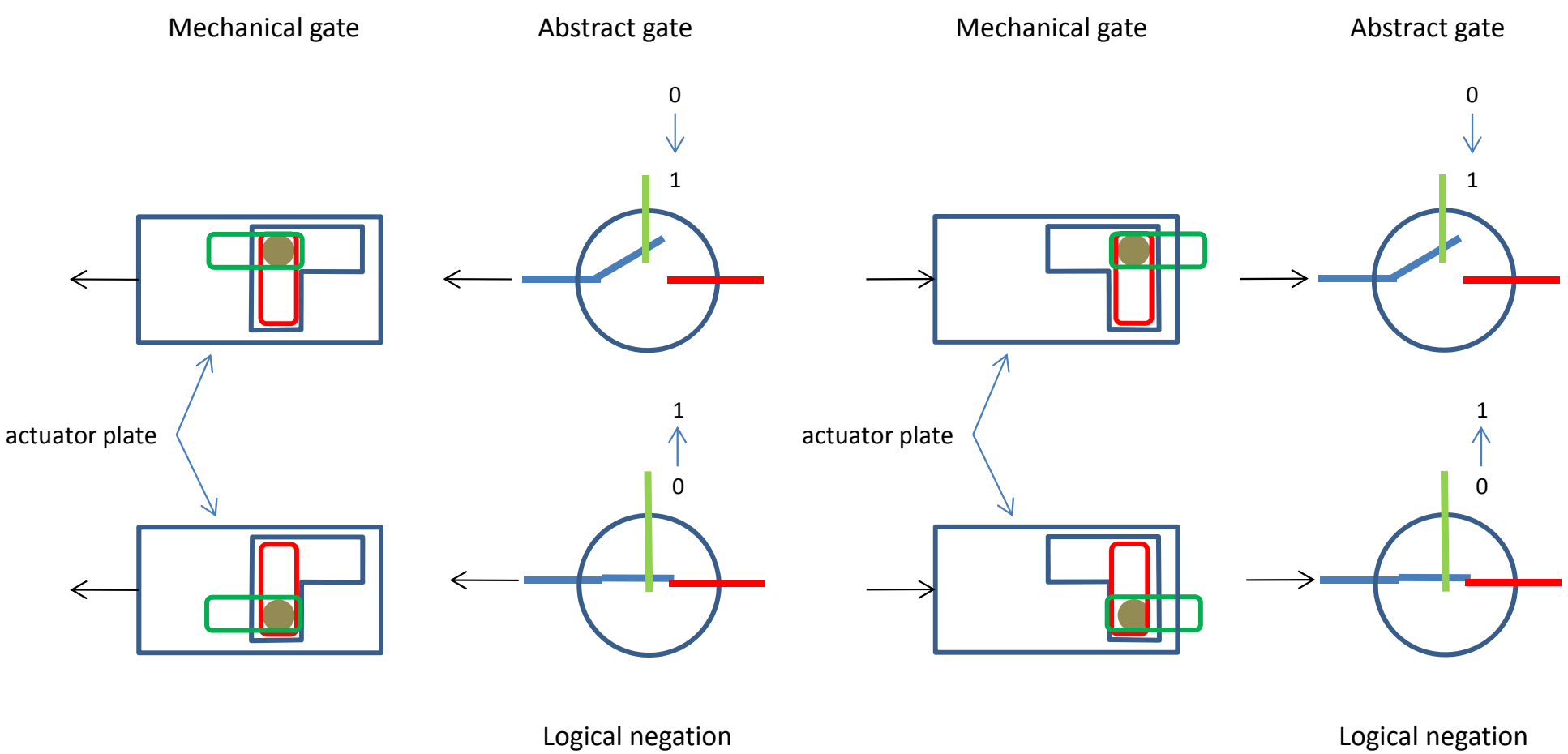

Fig. 7: Some variations of the elementary gate and Zuse's abstract notation for mechanical relays. The relays are drawn as switches. By convention, the drawing always shows the zero-position of the control bit. The arrows show the possible movements. The actuator plate can be pulled to the left (left side diagrams), or pushed to the right (right side diagrams). The initial position of the mechanical relay can be in the closed position (lower two diagrams). In that case the relay acts as a negation, since the output is the negation of the control bit.

Now everybody can start to build his/her Zuse mechanical computer. The basic element is the mechanical relay. More complex connections (like the relays with two actuated plates) can be designed, and the corresponding mechanics has to be built using plates and rods.

The main problem for building a complete computer is to interconnect all components. Notice that the control bit always moves orthogonally to the result bit. Each completed logical operation rotates the mechanical movement by 90 degrees. The next logical operation rotates the movement by 90 degrees, and so on. After four gates, we are back to the original direction of movement. This is why Zuse's cycling units used the four directions NESW. Within a machine cycle it is possible to execute four layers of logical computations. The logical gates can be simple, such as a negation, or complex, such as a gate with two actuated plates (half of an XOR). The clocking in the Z1 is such that the machine completes an addition in four engagements: in engagement IV the arguments are loaded. Engagements I and II compute partial sums and carries, and engagement III the final result.

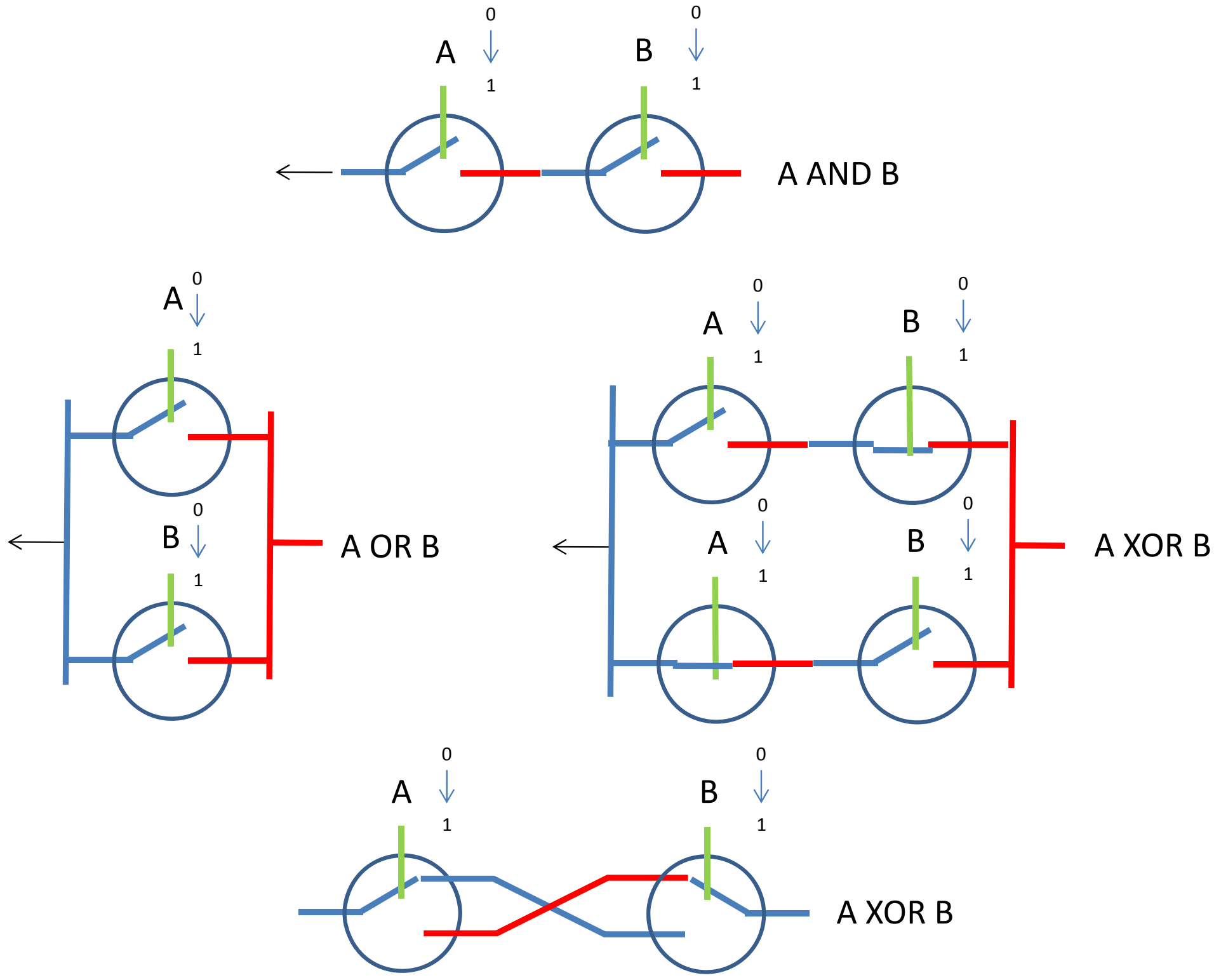

Fig. 8: Some logical gates built from mechanical relays. The lowest diagram, an XOR, can be built by using mechanical relays with two possible actuated plates, as shown in the diagrams. The mechanical equivalents are easy to design.

Result bits can be transferred to different horizontal levels than the level at which the input bits move. That is, rods can be also used to move bits "up" or "down" between the layers of the machine. We will see this later in the addition circuits.

At this point Fig. 5 should make more sense: the circles inside the different rectangles are exactly the circles of Zuse's abstract notation and pinpoint the position of logical gates. We can now abstract from the mechanics and discuss the Z1 from a more logical point of view.

## 6 The memory of the Z1

The memory of the Z1 was until now the best understood part of the Z1. It was described by Schweier and Saupe [4] in the 1990s. A very similar type of memory was used for the Z4, a relay computer finished by Konrad Zuse in 1945. The Z4 had a processor built with telephone relays, but the memory was mechanical, just like in the Z1. The mechanical memory of the Z4 is housed today at Deutsches Museum. Its operation has been simulated in a computer by a student assistant.

The main concept used in the Z1 was that a bit can be stored using a vertical pin which can be set in one of two possible positions. One position represents zero, the other position represents one. The diagram below shows how the memory bits can be set by moving them from one position to the other.

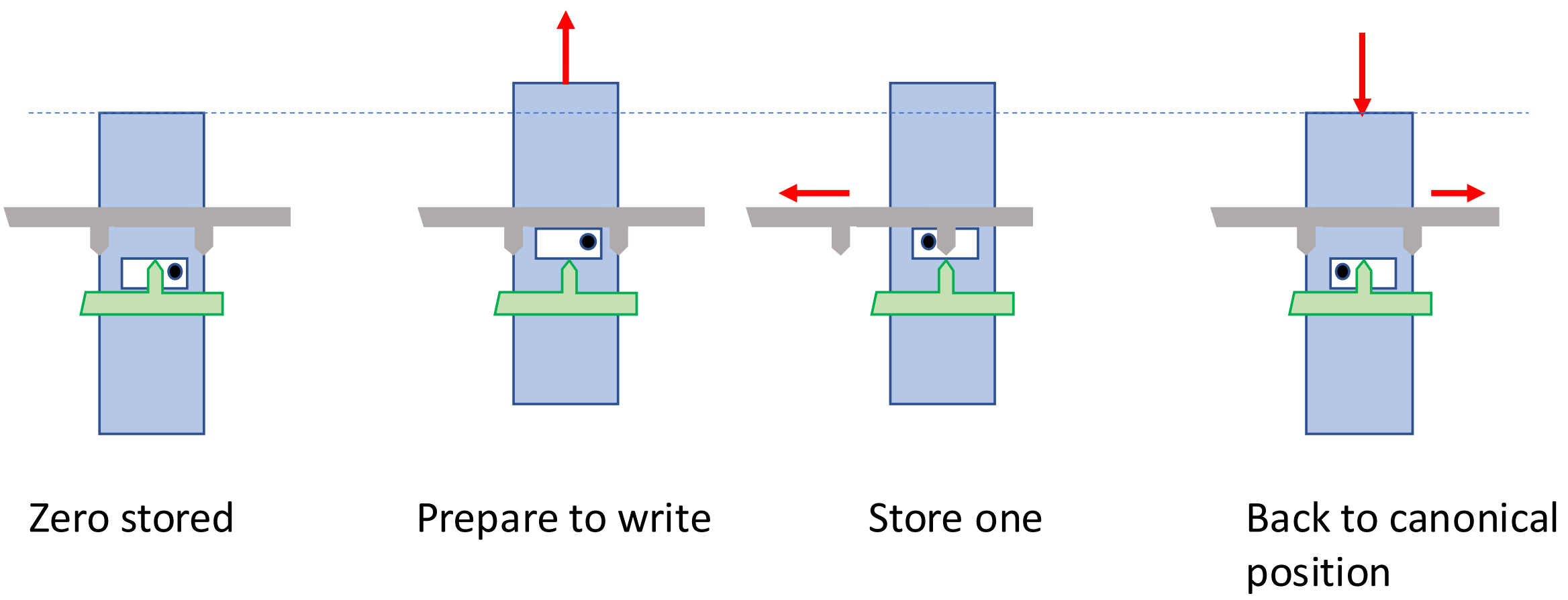

Figure 9: One mechanical bit in the memory. The pin can be stored in the zero or one position. Its position can be read.

Fig. 9 shows one stored bit (zero). In the first step, a control plate raises the pin. The horizontal actuated plate is pushed to the left. In the final step, the control plate moves down, and the canonical position of all elements is restored, but now the memory cell contains a one. Reading of the stored bit is done using a similar approach, but other control elements not shown here.

The memory words were addressed by decoding the 6 bits used for an address. Three bits selected one of 8 layers, the other three one of eight memory words. The decoding circuit for each layer was a classical binary tree of relays with three levels, as used in the Z3 (with a different number of levels).

We do not delve further into the structure of the mechanical memory. The details can be consulted in [4].

## 7 The addition unit of the Z1

The addition unit of the reconstructed Z1 differs from the type of addition unit described by Konrad Zuse in one document finished after the war. In that document [6], the binary digits are handled using OR, AND, and identity (NOT-XOR) logical gates. In the Z1 reconstruction, the addition unit uses two XORs and one AND computation.

During an addition, the first two computations performed are: a) the bitwise XOR of the two registers to be added, storing the result, b) the bitwise AND of the two registers to be added, storing the result (see first to rows of table below). The third step is the computation of the carry bits using the information from the AND and the XOR operation.

The example below shows how to add the binary numbers 10001 and 1, using the steps mentioned above. The successive bits, from the power $2^0$ to the power $2^4$, of each number are represented by $b_4, b_3, b_2, b_1, b_0$.

| | $b_4$ | $b_3$ | $b_2$ | $b_1$ | $b_0$ |
|---|---|---|---|---|---|
| Number 1 | 1 | 0 | 0 | 0 | 1 |
| Number 2 | 0 | 0 | 0 | 0 | 1 |
| XOR (propagator) | 1 | 0 | 1 | 1 | 0 |
| AND (generator) | 0 | 0 | 0 | 0 | 1 |
| Carries | 0 | 1 | 1 | 1 | 0 |
| Final result (XOR) | 1 | 1 | 0 | 0 | 0 |

Konrad Zuse used this so-called "anticipating carriage" in all his machines. Instead of propagating a carry through the different binary powers sequentially, the carry for all positions can be set in one step. The example above serves to illustrate the procedure. The first XOR is the partial result of the sum of the two registers without considering carriages. The AND computes the generation of carry bits: they are transported to the next bit to the left, but are further transported to the next binary position as long as there is a one in the result of the previous XOR computation. In the example, one carry is generated at the bit $b_0$ and is propagated to the bits $b_3$, $b_2$, and $b_1$. All carries are finally XORed with the result of the first XOR. A sequence of consecutive 1's from the first XOR operation operates like a kind of bandwagon for propagating AND-generated carries until the chain of 1's breaks.

The circuit shown in Fig. 10 is the addition circuit used in the reconstructed Z1. The diagram shows the addition of two bits stored in the $a$ and $b$ rods ($a$ could be the i-th bit of register Aa, and $b$ the corresponding bit of register Ab). The XOR and the AND computations are performed in parallel using the binary gates 1,2,3, and 4. The AND operates on gate 5, generating the carry bit $u_{i+1}$, while the XOR closes or leaves open the "chain" of XOR bits using gate 6. Gate 7 is an auxiliary gate for passing the XOR result to the upper level. Gates 8 and 9 compute the final XOR to complete the addition.

The movements of the different components are indicated by the arrows. All four cycle directions are used, that is, an addition takes one full cycle, from operand loading up to result generation. The result is passed to rod $e$, the i-th bit of the register Ae.

This addition circuit is located in layers 1, 2, and 3 of the addition block (see Fig. 13 further down). It is remarkable that Konrad Zuse, who had no formal training in binary logic, was working with anticipating carry. The ENIAC, the first large-scale electronic computer, propagated the carry sequentially from one decimal position in an accumulator to the next. The Harvard Mark I used anticipating decimal carry.

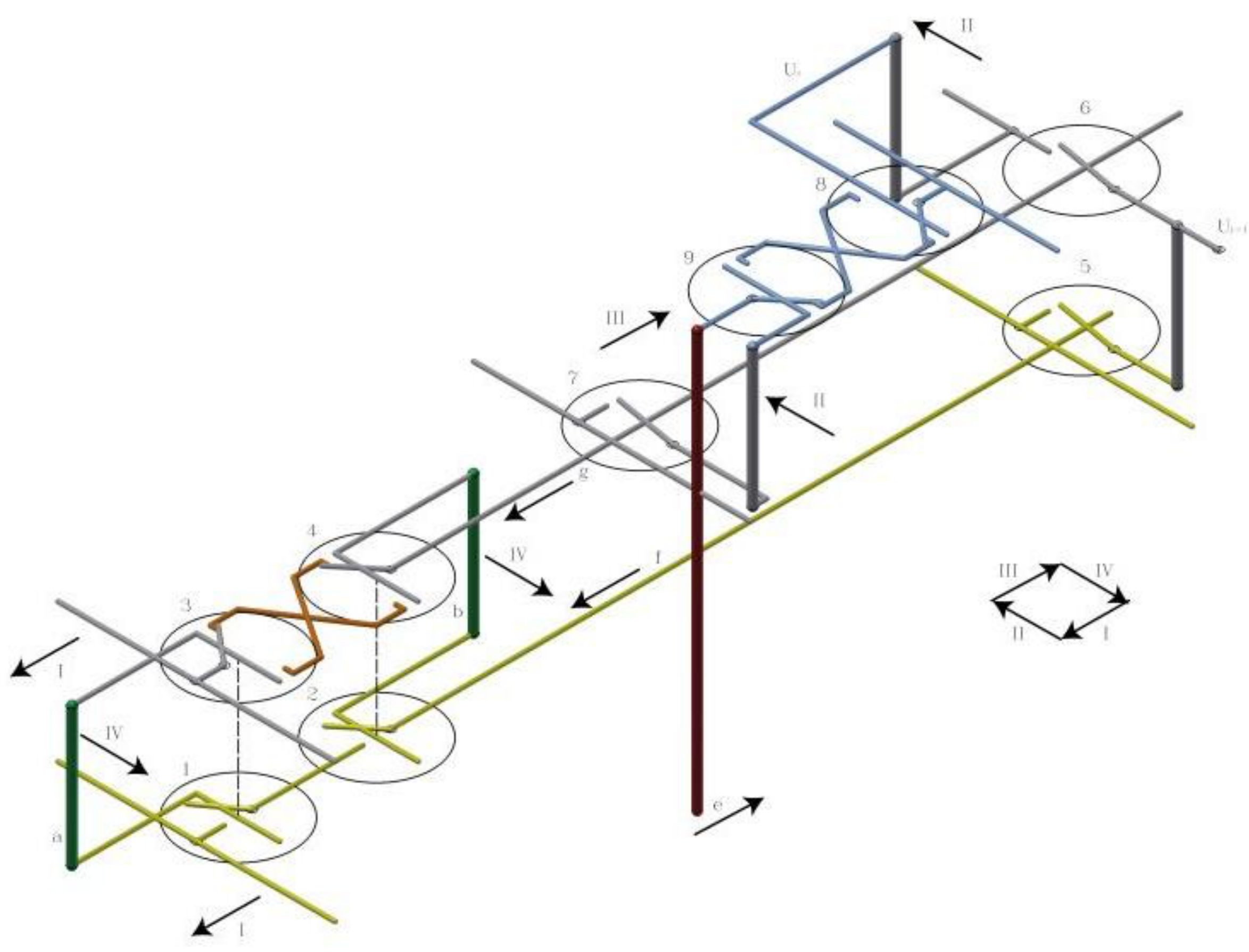

Fig. 10: Addition unit of the Z3. Computation runs from left to right. Bitwise AND and XOR are computed first (gates 1, 2, 3, 4). The carry bits are computed in engagement II (gates 5 and 6). In engagement III an XOR finishes the computation of the addition (gates 8 and 9).

## 7 The microinstructions of the Z1

Every operation in the Z1 is broken into a sequence of microinstructions. This is done through a kind of table of “criteria” consisting of 108 metallic plates placed in pairs, as shown in Fig. 11 (here we only see the top pair of plates, as seen from the top. The rest of the plates is situated below these two upper plates, across the 12 layers). The entries in the table (the metallic plates themselves) are ordered according to the values of ten bits:

- The Op0, Op1 and Op2 bits contain the binary opcode of the instruction
- The bits S0 and S1 are conditions bits, set by other parts of the machine. When S0=0, for example, an addition (S0=1) is transformed into a subtraction.
- The bits Ph0, Ph1, Ph2, Ph3, Ph4 are used to count the number of microcycles (or “phases”) in an instruction. Multiplication, for example, is executed in 20 phases and the five bits Ph0 to Ph4 advance from 0 to 19 during the operation.

The ten bits theoretically allow us to define up to 1024 different conditions or cases. An instruction can contain up to 32 phases. The ten bits (opcode, condition bits, and phase) push metallic pins (colored gray in Fig. 11) which stop the microcontrol plates from snapping to the left or right (each plate is attached to a spring, shown in the figure). Each microcontrol plate has a different configuration of teeth, which determines if the current position of the ten control pins can stop the plate from snapping or not. Each microcontrol plate has therefore an "address". When the ten control bits contains that address, the plate snaps to the right (the upper set of plates), or to the left (the lower set of plates in Fig. 11).

If a control plate moves to the right it presses on *one* of four condition bits (A, B, C, D). Each metallic plate is cut so that it presses on a specific contact A, B. C, or D according to the selected criterion. The four possibilities correspond to the four actions that can be triggered in each of the 12 layers in the processor.

Since the plates are mounted on top of each other, across the 12 layers of the machine, activating a control plate automatically selects the layer for the operation in the processor. One microoperation in the exponent unit can also be started parallel to a microoperation in the mantissa unit, since two plates can snap simultaneously: one to the left, another to the right. It is also possible for two plates to snap on two different layers of the right side (mantissa control), but mechanical constraints limit this kind of "parallelism".

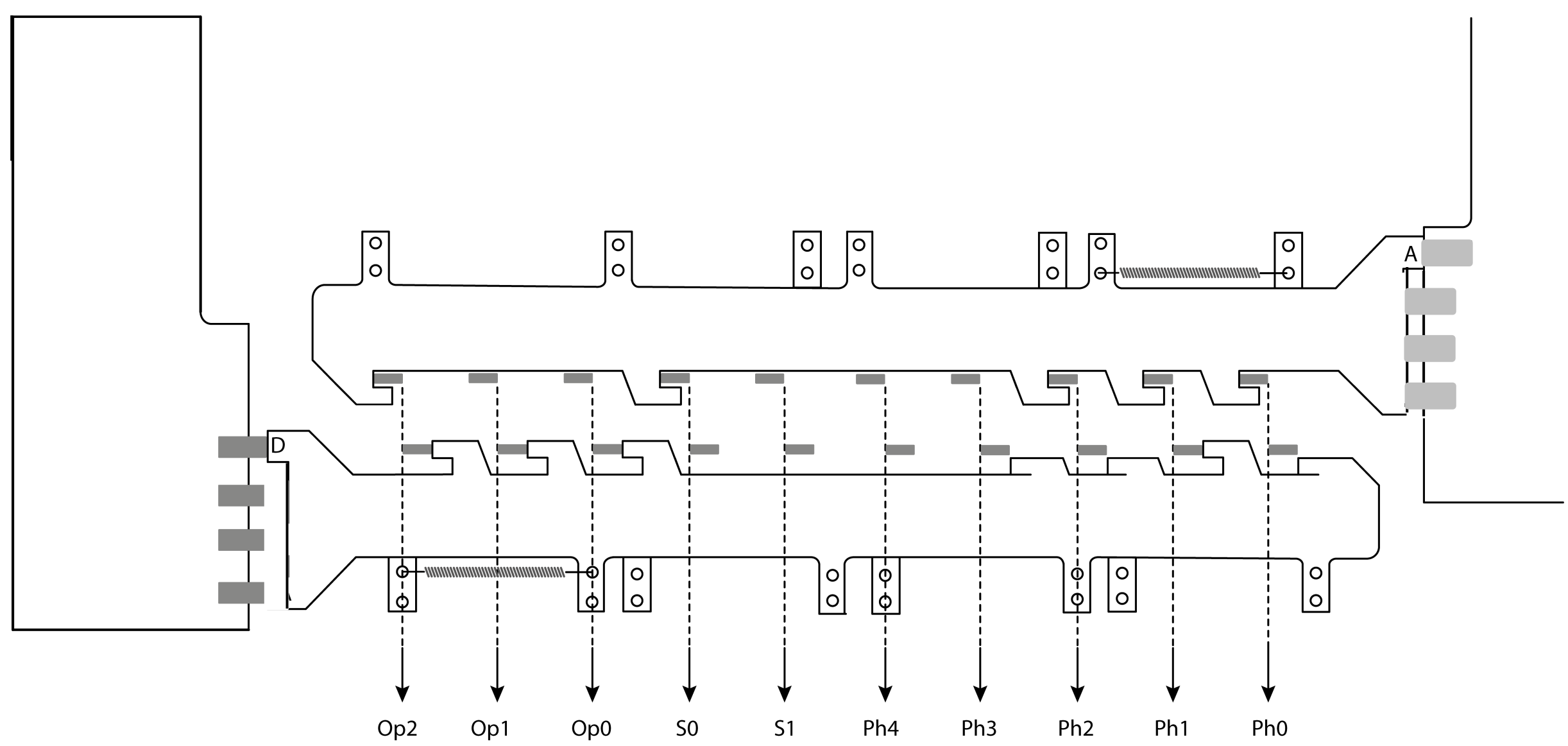

Fig. 11: The control plates. The teeth of each plate makes it stop, according to the position of the metallic pins (in grey), activated by the ten bits Op2 to Ph0. Springs make a plate snatch to the right (upper plate) or to the left (lower plate), when its "address" has been selected. Selecting a plate out from the 12-level stack of plates selects the layer for the next operation. The teeth A, B, C or D can be cut, so that only the necessary operation is selected by pressing on a specific pin in the microcontrol unit. In the diagram, the upper plate in layer 12 has moved to the right and presses on the pin A for layer 12.

Controlling the Z1 amounts thus to cutting the teeth of the metallic plates so that each one of them responds to a specific ten-bit combination, for acting on the left, or on the right side. The left side controls the exponent's half of the processor. The right side the mantissa half. The alternatives A, B, C or D are exclusive, that is, only one of them is selected by a microcontrol plate (by pushing it).

## 8 The Processor's Datapath

Fig. 12 shows the floating-point processor of the Z1. The processor has one block for handling the exponents (left side) and one for handling the mantissas (right side). The floating-point registers F and G consist of 7 bits for the exponent, and 17 bits for the normalized mantissa (numbered from $Be_0$ to $Be_{-16}$. The ALU has two more bits for $Be_{-17}$ and $Be_{-18}$, in order to produce correct rounding. It has also an extra bit $Be_1$, used in some algorithms.

The exponent-mantissa pair (Af,Bf) is FP register F and the pair (Ag,Bg) is FP register G. The signs of the arguments are handled externally, in a sign unit. The sign of a product or a division is computed in advance. The sign of an addition or subtraction is adjusted after the operation takes place.

In Fig. 12 we can see the registers F and G and their connections to the rest of the processor. The ALU (arithmetical logical unit) contains two FP registers: the pair (Aa,Ba) and the pair (Ab,Bb). These registers are the direct inputs to the ALUs. They have to be loaded and can retain partial results during several iterations, due to the feedback bus from the ALU-outputs Ae and Be.

In the Z1 the data buses are used in "three-state" mode, that is, many inputs can push on the same data line (which is a mechanical component). There is no need to isolate "electrically" the data lines from the inputs, since no electricity is in play. Since a zero input is represented by no movement of a mechanical part (no pushing), while a one represents a movement (pushing), there is no conflict between the parts. If two parts push on the same data line, the only important thing is that they act in step with the machine cycle (pushing only works in one direction).

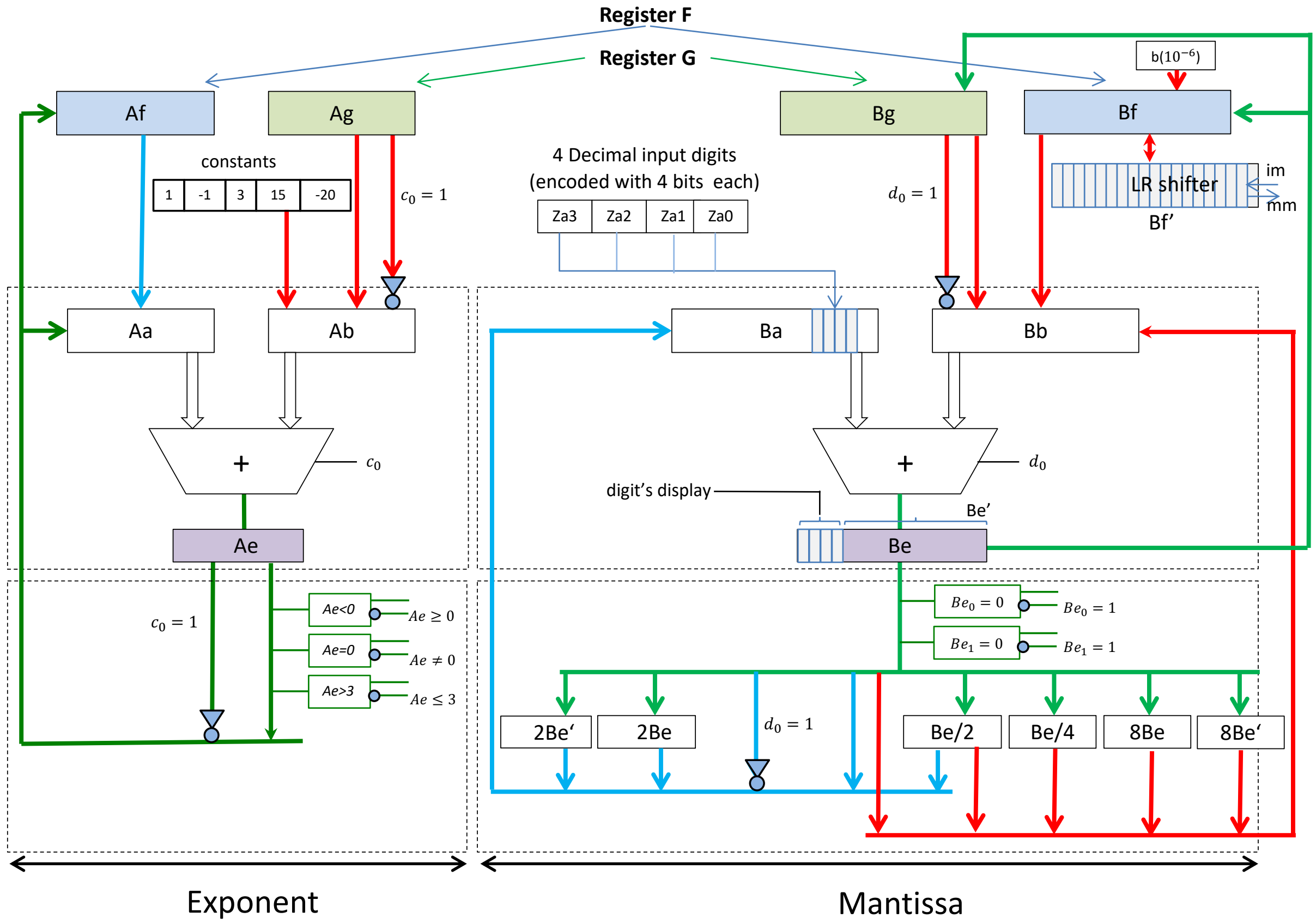

Fig. 12: The processor datapath in the Z1. The left part corresponds to the exponent's ALU and registers, the right side to the mantissa's. The results Ae and Be can be fed back to the temporary registers, or they can be negated, or shifted. The four bits representing a decimal digit (Za3, Za2, Za1, Za0) are copied to register Ba directly, one digit after another, using four bits. The decimal-binary conversion operates on this data.

The only registers visible to the programmer are (Af,Bf) and (Ag,Bg). They have no address: the first register loaded by a Load operation is (Af,Bf), the second register loaded afterwards is (Ag,Bg). Once two registers have been loaded, the arithmetical operations can be started. (Af,Bf) is also the result register for arithmetical operations. The second register can be loaded after an arithmetical operation and be the second argument for a new arithmetical operation. This scheme of register usage is similar as in the Z3. In the Z3, however, there is no register pair (Ag,Bg). The coordination between main and auxiliary registers is more complicated than in the Z1.

As can be seen from the processor datapath, the individual registers Aa, Ab, Ba, and Bb can be loaded with different kinds of data: values from other registers, constants $(+1, +1, 3, 15, -20)$, complement values of other registers, and the values coming back from the ALUs. The ALUs outputs can be complemented or shifted. A shift to the left by *n* places is represented by a box containing a multiplication with $2^n$; a shift to the right by *n* places by a division with $2^n$. These boxes are mechanical circuits containing the appropriate bit displacement or bit complementation. The result of the addition of the registers Ba and Bb, for example, is stored in Be and can be transformed in several ways: the result Be can be complemented (comp Be), can be shifted one or two places to the right (Be/2, Be/4), or it can be shifted one or three places to the left (2Be, 8Be). Every such computation is performed in a different layer of the mechanical stack of layers constituting the ALU.

When the Z1 needs to make a register negative, for example Ae, it complements its bits (represented by the negation gate for all bits) and sets the zero-carry bit in the ALU to +1 (in the diagram $c_0 = 1$). In the next pass through the ALU (using Aa), this is added as initial carry-bit to the complement, and this produces the negation of Ae (complement of Ae plus 1). The same trick is used to pass negative Bg to Bb, and negative Be to Ba. The other register added to the complement does not affect the intended result.

We see in the diagram that the ALU mantissa result, according to the active computation, is passed back to register Ba or Bb. Case selection (of the appropriate box or circuit in the diagram) is done by the microplates that activate the appropriate layer as selected by the microcontroller. The result Be can also go straight to the memory unit (the corresponding bus line is not shown in Fig. 12).

There are some condition boxes connected to Ae and Be. These are circuits that test individual bits of the ALU result, or conditions such as “Ae=0”.

The ALU performs an addition in every cycle. All registers Aa, Ab, Ba, and Bb are erased after an ALU computation, and can be reloaded with the feedback values or other registers.

Register Ba has a special use for the conversion of four decimal digits into binary. Each decimal digit entered through the mechanical panel is transformed into 4 bits. Groups of 4 bits are fed directly into register Ba (at position $2^{-15}$), which can advance the four bits by performing a multiplication with the factor 10, adding then the next digit to the partial result, multiplying again by ten, and so on. If we want to transform the number 8743 from decimal to binary, for example, the digit 8 is entered first (in binary) and is multiplied by ten. Then 7 (in binary) is added to the result, and the new sum (87) is multiplied by ten. The result (870) is then added to 4, and so on. This yields a simple algorithm for the conversion of decimal input to a binary number. During the process the exponent half of the processor is adjusting the exponent of the final floating-point result (therefore the constant 15 in the exponent ALU, which corresponds to $2^{15}$ (see decimal-binary conversion algorithm further down).

Register Be is also used for the output of a decimal number. Roughly speaking, if the contents of Ae is the number 0.8754, in decimal, we can extract the successive digits multiplying Ae by ten, to obtain first 8.754. The 8 is shown in the mechanical output display and is erased from Ae (using the top four bits of Ae). A new multiplication by 10 of Be' (as Be with the previous erasure of the first four bits is called) is done, obtaining 7.54 in Be. We display the 7 and continue iteratively with the rest.

The circuits for computing 2Be and 8Be are useful: their sum in the ALU produces the desired multiplication by 10. The same is true for Be', that is, the mantissa with the erased top four bits.

## 9 Operational block structure

Fig. 13 shows the main elements of the flow of control in the Z1. First, an instruction is read from the tape reader. If it is a load instruction, the memory unit is called upon to send the contents of the given address to the memory-processor interface. If it is a store instruction, register (Af,Bf) is stored in memory into the corresponding memory address. In fact, the registers (Af,Bf) and (Ag,Bg) are connected directly to the memory-processor interface. The memory can only send data to and read data from the processor through both registers. The first time data is sent to the processor, it is loaded to (Af,Bf). The second time, it is loaded to (Ag,Bg). Results are stored in (Af,Bf), and (Ag,Bg) is cleared.

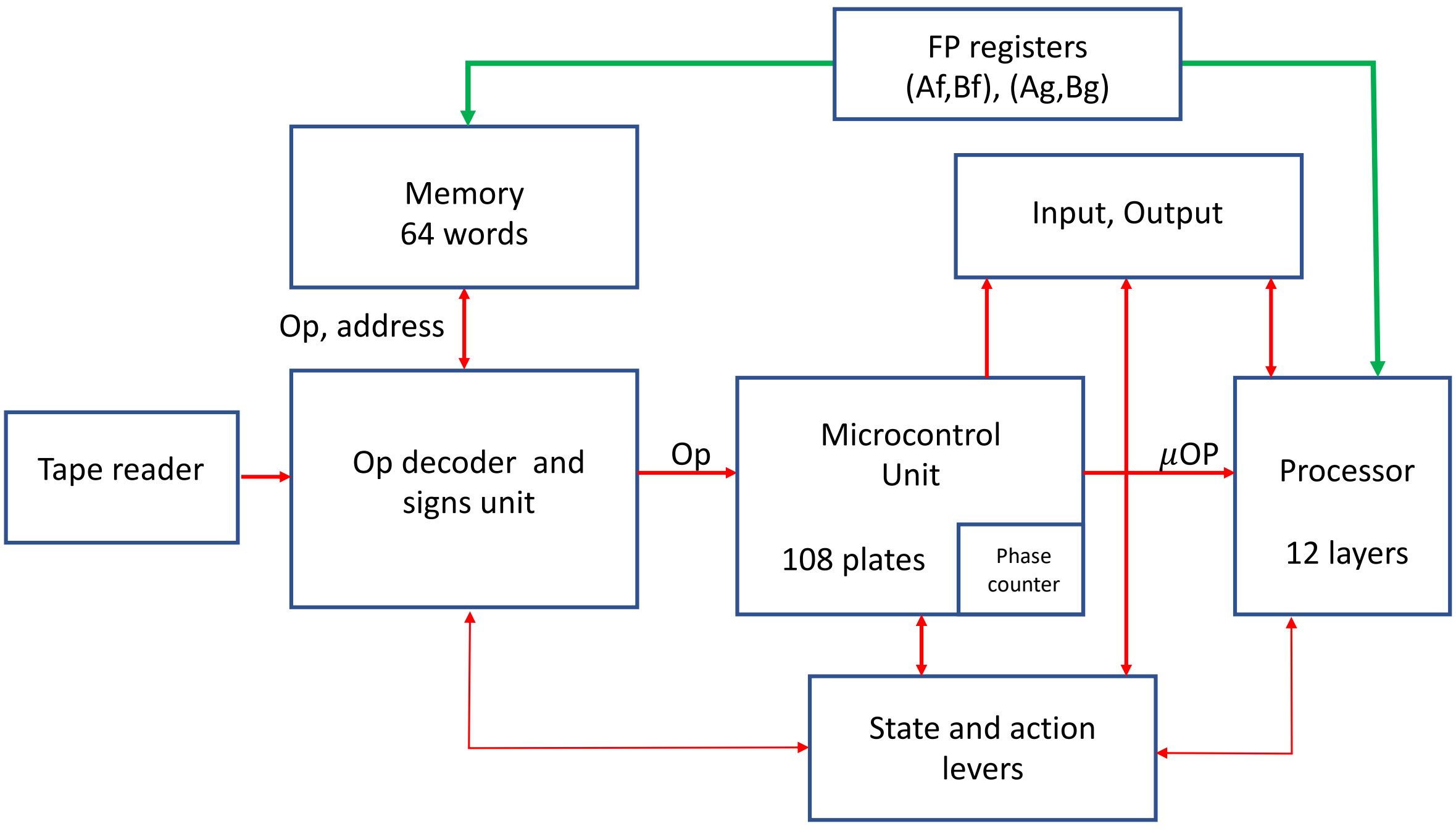

Fig. 13: The control flow when executing a program in the Z1

The Op decoder and signs unit decides if the instruction involves memory, in which case the procedure explained above is used. The signs unit keeps track of the signs of the numbers loaded in (AG,Bg) and (Af,Bf) and precomputes the sign of the result of an operation. In the case of a multiplication or division, the sign computed is final, using the rule of signs. In the case of addition or subtraction, the preliminary sign can change according to the result of the algorithm (see section x).

If the instruction does not involve memory, the microcontrol unit is called upon, for the specific operation. This unit has a kind of table of conditions that result in actions. The condictions are given by the opcode of the instruction, some flags, and the phase (cycle) of the operation. A subtraction can take 7 cycles and the "phase counter" advances from phase 0 to phase 6, triggering each time another microinstruction from the table.

Inside the Z1, the microinstructions are represented by plates that snap to the left or right, in order to trigger a microoperation in the exponent part of the processor, or in the mantissa part. There are some state and action levers that can be set or cleared according to what happens in the processor, or possible user interaction. The lever u8 for example, notifies the processor that the user has finished entering a decimal number through the decimal board, and that input can be transformed into binary. The lever (Ae) orders the processor to copy register Ae to register Aa. It is a very useful microoperation.

The input (the decimal keyboard) and the output (decimal display) are connected directly to the processor. Their interplay is regulated by the microcontrol unit.

## 10 Layered construction

Fig. 14 shows the spatial distribution of some different elements of the processor datapath for the mantissa part. All the shifters have been allocated in different layers of the twelve constituting the leftmost module of the machine. The registers Bf and Bg come from the right side, directly from memory (layers 5 and 7). The result Be is fed back to memory crossing through level 8. The bits of the registers Ba, Bb, and Be are stored in vertical rods (only one bit is shown in this cross-section of the processor). The ALU is distributed in two mechanical stacks. Level 1 and 2 compute the AND and XOR of the individual bits in Ba and Bb. The results are passed to the right, where the carry-bits and the final XOR is computed and stored in Be. The result Be can go back to be stored in memory, or can be shifted in all the different ways shown, being fed back to Ba or Bb, as desired. Some circuits seem redundant (there are two ways of loading Be into Ba, for example), but they represent alternatives. Level 12 loads Be into Ba unconditionally, level 9 only if the exponent Ae is zero. The boxes marked green in the diagram are empty layers, where no computation takes place and mechanical components can pass through. The box around the bars Bf and Bf' contains the shifter for Bf needed for multiplication (where each bit of Bf is read one after the other, starting from the lowest binary power).

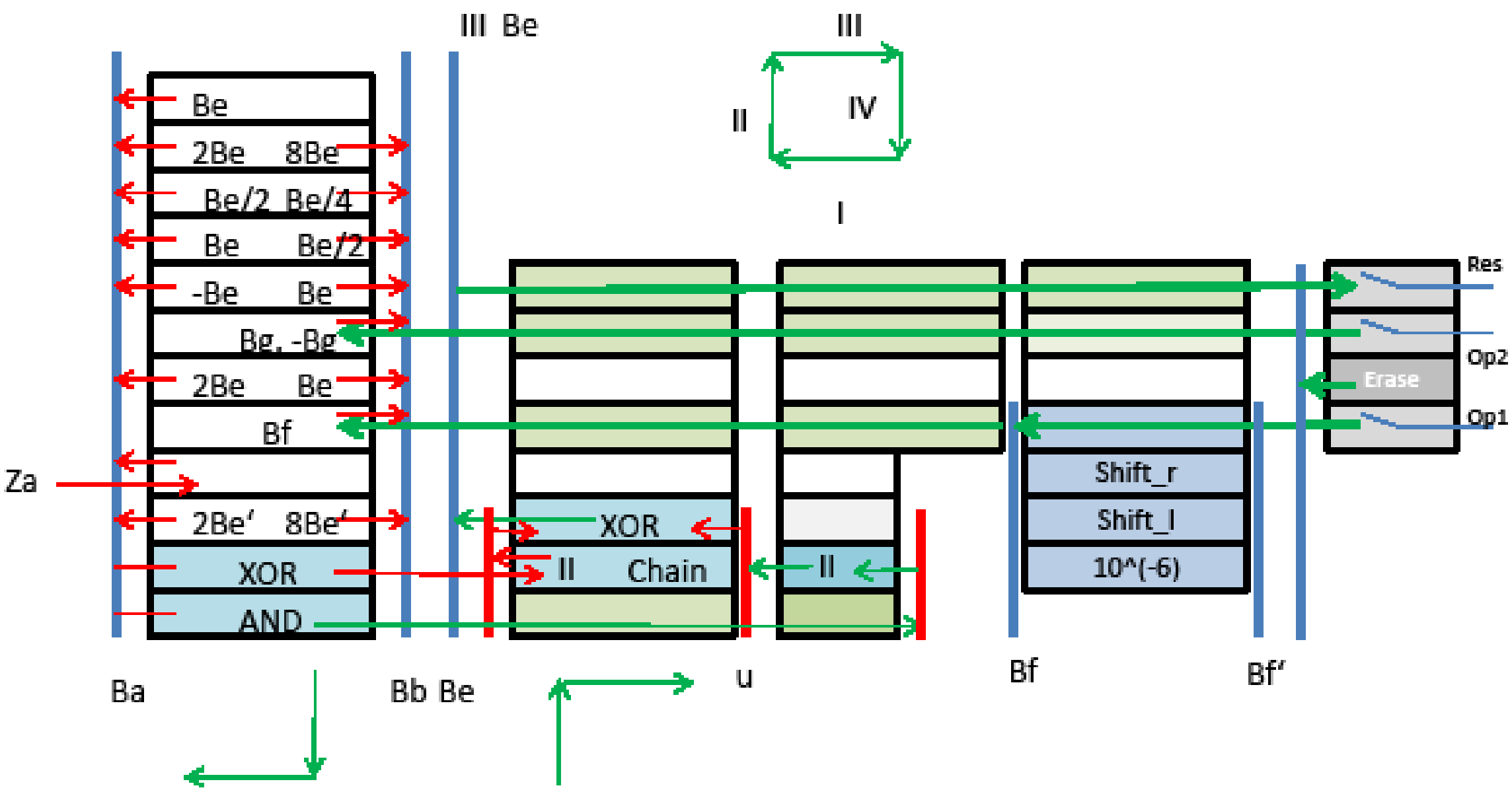

Fig. 14: The layered spatial distribution of operations in the processor. The shifters for Be are on the left stack. The addition unit is distributed between the three leftmost stacks. The shifters for Bf are in the right stack, and the binary equivalent of $10^{-6}$ too. The result goes to memory through the line labeled Res on the right. The two registers Bf and Bg arrive from memory as First (Op1) or second operand (Op2).

Now you can picture the computational stream in this machine: data flows from the registers F and G into the machine, filling the A and B register pairs. A single addition or sequences of additions/subtractions (for multiplication or for division) are performed. Partial results are recycled in the A and B registers until the result is complete. The final result is then loaded in register F and a new computation can be started again.

## 9 The arithmetical instructions

As explained above, the Z1 could perform the four basic arithmetical operations. In the tables discussed below, the convention has been used of representing a binary one with the letter "L". The tables display the sequence of microinstructions needed for each operation and how it affects the dataflow between the registers in the processor. One table summarizes addition and subtraction (using two's complement), one table summarizes multiplication, and one table is for division. There is also one table for each of the two I/O operations: decimal-binary and binary-decimal conversion. The tables are divided into Part A for the exponent, and part B for the mantissa. The registers Aa, Ab, Ba, and Bb are loaded as shown in each row of a table. The phase of the operation is given by the column labeled "Ph". Conditions can trigger or inhibit an operation from starting. If a row is executed, condition bits can be set, or the next phase (Ph) can be computed by the incrementer.

In the table the "Level" refers to one of the 12 levels of the mechanics (as shown in Fig. 14). The word No. (number) is the successive numbers of the different microcontrol plates. "Criterion" refers to the Opcode (three bits), two flags, and the phase. Ph is the "phase" of the instruction, that is, the successive microinstructions, from 0 to 6 for addition, for example. Sometimes the criterion is combined with a condition to trigger the microcode-plate, and some state bits or parts are "operated" upon. For example, in Fig. 15, the line before the last one is an "alignment" for the mantissa (normalization). In the exponent part A this happens in layer 6, activated by microcontrol plate 6, for the addition, in phase 6. If $Be_0 = 0$, Ae is loaded to Aa and -1 to Ab. The ALU produces $Ae - 1$, and so on. For the mantissa, level 6 is working, activated by plate 12, and so on.

**Addition/Subtraction**

The table of microinstructions below covers both the case of an addition of numbers, as well as a subtraction. The main problem for both operations is to scale the two numbers to be added/subtracted so that the binary exponent is the same. Assume that the numbers $m_1 \times 2^a$ and $m_2 \times 2^b$ are to be added. If $a = b$ the two mantissas can be added immediately. If $a > b$ then the smaller number is rewritten as $m_2 \times 2^{b-a} \times 2^a$. The first multiplication is equivalent to shifting the mantissa $m_2$ by (*a*-*b*) places to the right (making the mantissa smaller). Let us call $m_2' = m_2 \times 2^{b-a}$. The two mantissas to be added are now $m_1$ and $m_2'$. The common binary exponent is $2^a$. A similar procedure is used in case that $a < b$.

The table shows the phase (Ph) progressing from 0 to 6. The opcode is that of the addition instruction. The "level" refers to the layer in the processor (from 1 to 12). The left part of the table refers to the exponent and the right part to the mantissa.

| Part A (Exponent) | | | | | | | | | | | | | | Notification->B | Part B (Mantissa) | | | | | | | | | | | |
|---|---|---|---|---|---|---|---|---|---|---|---|---|---|---|---|---|---|---|---|---|---|---|---|---|---|---|
| Block 17 (left Part) | | | | | | | | | | Block 15a | | | | | Block 17 (right part) | | | | | | | | Block 15b | | | |
| Operating sequence | | Level | Criterion | | | | | Cond. | to operate | Condition | to operate | Aa | Ab | | Level | Criterion | | | | | Conditio n | to operate | Condition | to operate | Ba | Bb |
| Symb | Description | | No. | Op | S0 | S1 | Ph | | | | | | | | | No. | Op | S0 | S1 | Ph | | | | | | |
| ADDITION (Σ) | Difference of the Exponents Δα | 5 | 1 | Σ | | | 0 | | Ph | | | Af | | | | | | | | | | | | | | |
| | | 7 | 1 | Σ | | | 0 | | | | | | -Ag | | | | | | | | | | | | | |
| | S1 ? | 4 | 2 | Σ | | | 1 | | Ph, (Ae) | Ae ≥ 0 | S1 | Ae | | | | | | | | | | | | | | |
| | \|Δα\| | 3 | 3 | Σ | | 0 | 2 | | Ph | | | -Ae | | | 5 | 3 | Σ | | 0 | 2 | | | | | | Bf |
| | | 4 | 4 | Σ | | L | 2 | | (Ae) | | | Ae | | | 7 | 4 | Σ | | L | 2 | | Ph | | | | Bg |
| | Align → | 9 | 5 | Σ | | | 3 | | Ae | | | | | | | | | | | | | | | | | |
| | | | | | | | | | | Ae = 0 | Ph | | | B | 9 | (5) | Σ | | | 3 | | A → | (Ae = 0) | | Be | |
| | | | | | | | | | | Ae ≠ 0 | | | -1 | B | 9 | (5) | Σ | | | 3 | | A → | (Ae ≠ 0) | | | ½ Be |
| | Actual Addition (Σ) | | | | | | | | | | | | | | 8 | 6 | Σ | 0 | | 4 | | | | | -Be | |
| | | | | | | | | | | | | | | | 12 | 7 | Σ | L | | 4 | | | | | Be | |
| | | 5 | 8 | Σ | | L | 4 | | Ph | | | Af | | | 5 | 8 | Σ | | L | 4 | | | | | | Bf |
| | | 7 | 9 | Σ | | 0 | 4 | | Ph | | | | Ag | | 7 | 9 | Σ | | 0 | 4 | | (Ph) | | | | Bg |
| | Align → | 2 | 10 | Σ | L | | 5 | | Lz, Ae | | | Ae | | | 9 | 10 | Σ | L | | 5 | | | $Be_{+1} = 0$ | | Be | |
| | | 8 | 10 | Σ | L | | 5 | | | $Be_{+1} = L$ | | Ae | +1 | | 9 | 10 | Σ | L | | 5 | | | $Be_{+1} = L$ | | | ½ Be |
| | Complement by real Substraction | 8 | 11 | Σ | 0 | | 5 | | Ph, Ae | $Be_{+1} = L$ | S3 | Ae | | | 8 | 11 | Σ | 0 | | 5 | | | $Be_{+1} = L$ | | -Be | |
| | | | | | | | | | | | | Ae | | | 8 | 11 | Σ | 0 | | 5 | | | $Be_{+1} = 0$ | | | Be |
| | Alignment by real Substraction | 6 | 12 | Σ | 0 | | 6 | | Ae | $Be_0 = 0$ | | Ae | -1 | | 6 | 12 | Σ | 0 | | 6 | | | $Be_0 = 0$ | | 2Be | |
| | | | | | | | | | | $Be_0 = L$ | Lz | Ae | | | | | | | | | | | $Be_0 = L$ | | | Be |

Fig. 15: The microinstructions for addition and subtraction. An addition is finished in 5 cycles, a subtraction in 6.

When the microinstruction Ph is "operated upon", this means that Ph advances by one, to the next instruction step. The registers Aa and Ab are loaded as specified, as well as the registers Ba and Bb (refer to the diagram of the datapath). The phases 0 to 5 are used for the addition (S0 has been set to L). If S0 has been set to zero, a subtraction is performed in Phase 6.

The column "No." in the table refers to the number of the plate used in the microinstruction unit, to select a specific step.

- In Ph=0, the exponents of the two arguments are subtracted. If the result is positive or zero (Ae>=0) then S1 is set to one.
- In Ph=2, if the result was negative (S1=0), it is negated so that we effectively compute the absolute difference Δ of exponents.
- In Ph=3, Δ is used as a counter in order to shift the smaller mantissa Ae places to the right (in Ph=3). That phase is repeated until the counter is zero.

- In Ph=4, the exponent and mantissa of the larger argument are loaded to the register pair (Aa,Ba).
- In Ph=5, the addition or subtraction is performed (condition bit S0 is tested, and if one, an addition of the mantissas is performed, otherwise a subtraction). If the addition was performed, the algorithm finishes with Lz (clear and copy result to register (Af,Bf)).
  If it was a subtraction, then the result in the mantissa side could be negative. In that case, it is negated so that the mantissa of the result is positive. This change of sign is written down in the condition bit S3, in order to make the necessary adjustment to the sign of the final result.
- In Ph=6, the result of the subtraction is normalized, shifting the mantissa to the left, if needed, until the leading bit is 1.

The sign unit near the tape reader (see Fig. 5, Block 16) computes in advance the sign of the result and the type of operation. If we assume that the mantissas *x* and *y* are positive, then we have the following four cases for addition and subtraction (after having distributed the signs. For example, 2-(-3) is 2+3 after distributing signs). We call the result *z*:

1) $z = +x +y$
2) $z= +x -y$
3) $z= -x +y=-(x-y)$
4) $z= -x -y=-(x+y)$

The cases (1) and (4) can be handled with an addition in the ALU. In case (1) the result will be positive. In case (4) it will be negative. Cases (2) and (3) require a subtraction. The sign of the subtraction is computed in phase 5 (Fig. 15), and is adjusted according to the case 2 or 3 that we are handling.

Summarizing, an addition runs in the following steps:

- Determine difference $\Delta a$ of exponents in the exponent unit,
- Select largest exponent,
- Shift mantissa of the smaller number $\Delta a$ times to the left,
- Add the mantissas,
- Normalize the result,
- The sign of the result is the sign of both arguments.

A subtraction runs in the following steps:

- Determine difference $\Delta a$ of exponents in the exponent unit,
- Select largest exponent,
- Shift mantissa of the smaller number $\Delta a$ times to the left,
- Subtract the mantissas,
- Normalize the result,
- The sign of the result is the sign of the largest mantissa (in absolute value).

The final sign of the result is negotiated with the sign unit, which has a preliminary sign for the operation.

## Multiplication

A multiplication proceeds as follows:

- In Ph=0, the exponents of the two numbers are added in phase 0 (criterion 21, exponent part). At the same time the shifter is activated using lever ‘’R4.
- For the multiplication, the mantissa Bf is copied to the shifter Bf’. In the Phases 1 to 17, every bit of the mantissa Bf’ is shifted out to the right, that is, starting from the lowest power, all the way to the highest binary power in the mantissa (from $Bf_{-16}$ to $Bf_0$). In each phase, the bit *mm* contains the shifted-out bit at position -16. If the shifted-out bit is 1, then Bg is added to the partial result (which has been shifted previously one position to the right), otherwise zero is added. This algorithm computes therefore the final result

$$Be = Bf_0 \cdot Bg + \frac{1}{2}\left(Bf_{-1} \cdot Bg + \frac{1}{2}\left(Bf_{-2} \cdot Bg \ldots + \frac{1}{2}(Bf_{-16} \cdot Bg) \ldots\right)\right)$$

$$Be = Bf_0 \cdot 2^0 \cdot Bg + Bf_{-1} \cdot 2^{-1} \cdot Bg + \cdots + Bf_{-16} \cdot 2^{-16} \cdot Bg$$

- In Ph=18, if the mantissa after the multiplication is larger or equal than 2, the result is normalized by shifting the result one position to the right.
- In Ph=19, the final result is put in the data bus.

<table>
<tr><th colspan="10">Part A (Exponent)</th><th colspan="8">Part B (Mantissa)</th></tr>
<tr><th colspan="7">Block 17 (left side)</th><th colspan="3">Block 15a</th><th colspan="5">Block 17 (right side)</th><th colspan="3">Block 15b</th></tr>
<tr><th colspan="2">Operation</th><th colspan="4">Criterion</th><th rowspan="2">to set</th><th rowspan="2">Condition</th><th rowspan="2">Aa</th><th rowspan="2">Ab</th><th colspan="4">Criterion</th><th rowspan="2">to set</th><th rowspan="2">Condition</th><th rowspan="2">Ba</th><th rowspan="2">Bb</th></tr>
<tr><th>Symb</th><th>Description</th><th>No.</th><th>S0</th><th>S1</th><th>Ph</th><th>No.</th><th>S0</th><th>S1</th><th>Ph</th></tr>
<tr><td rowspan="7">Multiplication (×)</td><td rowspan="2">Sum of the Exponents Δα</td><td>21</td><td>0</td><td>0</td><td>0</td><td>Ph</td><td></td><td>Af</td><td></td><td></td><td></td><td></td><td></td><td></td><td></td><td></td><td></td></tr>
<tr><td>21</td><td>0</td><td>0</td><td>0</td><td>"R4.</td><td></td><td></td><td>Ag</td><td></td><td></td><td></td><td></td><td></td><td></td><td></td><td></td></tr>
<tr><td rowspan="2">actual Multiplication (×)</td><td>24</td><td></td><td></td><td>1<br>⋮<br>17</td><td>Ph, "R4, Ae</td><td></td><td></td><td></td><td>24</td><td></td><td></td><td>1<br>to<br>17</td><td></td><td></td><td>½ Be</td><td></td></tr>
<tr><td></td><td></td><td></td><td></td><td></td><td></td><td></td><td></td><td>24</td><td></td><td></td><td>1<br>to<br>17</td><td></td><td>mm = L</td><td></td><td>Bg</td></tr>
<tr><td rowspan="2">Align →</td><td>26</td><td></td><td></td><td>18</td><td>Ph, Ae</td><td></td><td></td><td></td><td>26</td><td></td><td></td><td>18</td><td></td><td>$Be_{+1} = 0$</td><td>Be</td><td></td></tr>
<tr><td>26</td><td></td><td></td><td>18</td><td></td><td>$Be_{+1} = L$</td><td></td><td>+1</td><td>26</td><td></td><td></td><td>18</td><td></td><td>$Be_{+1} = L$</td><td></td><td>½ Be</td></tr>
<tr><td>Finish</td><td>27</td><td></td><td></td><td>19</td><td>Ae</td><td></td><td>Ae</td><td></td><td>27</td><td></td><td></td><td>19</td><td></td><td></td><td>Be</td><td></td></tr>
</table>

Fig. 16: The microinstructions for multiplication. The multiplier-mantissa Bf is stored in a shift register (shift right). The multiplicand-mantissa is stored in register Bg.

Fig. 17 is the complete table for the multiplication instruction. It includes additional steps which are used for the decimal-binary and binary-decimal conversion. What is needed in those instructions is a multiplication with the number $10^{-6}$ which is loaded in Ph=0. In the exponent part we load the exponent of $10^{-6}$, denoted by $a(10^{-6})$, which is -20, and in the mantissa part, the mantissa of $10^{-6}$, denoted by $b(10^{-6})$ is moved to Bf. Then the multiplication is started. The reader can imagine that in that case, the multiplication instruction is “called” by the I/O instructions. The rest of the multiplication proceeds as usual, and at the end of the call some additional cycles finish the I/O operation.

| Part A (Exponent) | | | | | | | | | | | | | | Notification->B | Part B (Mantissa) | | | | | | | | | | | |
|---|---|---|---|---|---|---|---|---|---|---|---|---|---|---|---|---|---|---|---|---|---|---|---|---|---|---|
| Block 17 (left Part) | | | | | | | | | | Block 15a | | | | | Block 17 (right part) | | | | | | | | Block 15b | | | |
| Operating sequence | | Level | Criterion | | | | | Cond. | to operate | Condition | to operate | Aa | Ab | | Level | Criterion | | | | | Conditio n | to operate | Condition | to operate | Ba | Bb |
| Symb | Description | | No. | Op | S0 | S1 | Ph | | | | | | | | | No. | Op | S0 | S1 | Ph | | | | | | |
| Multiplication (×) | Sum of the Exponents Δα | 5 | 21 | × | 0 | 0 | 0 | | Ph | | | Af | | | | | | | | | | | | | | |
| | | 7 | 21 | × | 0 | 0 | 0 | | "R4. | | | | Ag | | | | | | | | | | | | | |
| | | 5 | 22 | × | L 0 | 0 L | 0 | | Ph, Ae, $b(10^{-6})$ | | | | $a(10^{-6})$ | | | | | | | | | | | | | |
| | actual Multiplication (×) | 7 | -24 | × | | | 1 ⋮ 17 | | Ph, "R4, Ae | | | Ae | | | 10 | 24 | × | | | 1 ⋮ 17 | | | | | ½ Be | |
| | | | | | | | | | | | | | | | 7 | 24 ↓ | × | | | 1 ⋮ 17 | | | mm = L | | | Bg |
| | | | | | | | | | | | | | | | is created in [8] | | | | | | | | | | | |
| | Align → | 7 | 26 | × | | | 18 | | Ph, Ae | | | Ae | | | 9 | 26 | × | | | 18 | | | $Be_{+1} = 0$ | | Be | |
| | | 8 | 26 | × | | | 18 | | | $Be_{+1} = L$ | | Ae | 1 | | 9 | 26 | × | | | 18 | | | $Be_{+1} = L$ | | | ½ Be |
| | Finish | 4 | 27 | × | | | 19 | | Ae | | | Ae | | | 12 | 27 | × | | | 19 | | | | | Be | |
| | | 2 | 28 | × | | 0 | 19 | | Lz, (Ae) | | | | | | | | | | | | | | | | | |
| | Setting ↘ | | | | | | | | | | | | | | 1 | 29 | × | | L | 19 | | ↘ , Ph1 | | | | |
| | Setting $b(10^{-6})$ | | | | | | | | | | | | | | 2 | 30 | × | L | | 0 ⋮ 5 | | u5 | | | | |
| | | | | | | | | | | | | | | | 2 | 31 | × | | L | 0 ⋮ 5 | | d4 | | | | |

Fig. 17: The microinstructions for multiplication with all the information and with the additional instructions for the I/O operations.

## Division

The division algorithm takes 21 cycles and is based on so-called “non-restoring floating-point division”. The bits of the quotient are computed one by one, starting from the highest order bit and moving to the lower-order bits successively.

First the difference of the exponents is computed in cycle 0, and then the division of the mantissas is executed. The divisor mantissa has been stored in Register Bg and the dividend mantissa in register Bf. The remainder is initialized to Bf in cycle 0. In each cycle thereafter, the divisor is subtracted from the remainder. If the result is positive, the corresponding bit in the mantissa of the result is set to one. If the result is negative, the bit in the result mantissa is set to zero. The result bits are computed one after the other, from bit zero to bit -16. There is a shifter in the Z1 for setting the bits of register Bf one after the other, as needed.
If the remainder becomes negative, there are two possible strategies for continuing. In “restoring division” the divisor D is added back to the remainder (R-D), in order to come back to the positive remainder R. Then the remainder is shifted one position to the left (which is equivalent to shifting the divisor to the right) and the algorithm continues. In “non-restoring division”, the remainder (R-D) is shifted one position to the left and then the divisor D is added. Since in the previous step (R-D) was negative, the shift to the left transforms this quantity in (2R-2D). If we now add the divisor, we obtain (2R-D), which is the subtraction of D from the shifted positive R, as we would like to have in the next step of the division algorithm. The algorithm can continue in this manner until the remainder becomes positive, and then we continue by subtracting again the divisor D. In the table below $u_{+2}$ refers to the carry bit for the binary power at position 2. If this bit is set, the result of the addition was negative (in two’s-complement arithmetic).
Non-restoring division is a very elegant way of computing the quotient of two floating-point mantissas, since the restoring step (one additional cycle) is avoided.

| Part A (Exponent) | | | | | | | | Part B (Mantissa) | | | | |
|---|---|---|---|---|---|---|---|---|---|---|---|---|
| Block 17 (left side) | | | | | Block 15a | | | Block 17 (right side) | | Block 15b | | |
| Operation | | Criterion | | to operate | Condition | Aa | Ab | Criterion | | Condition | Ba | Bb |
| Symb | Description | No. | Ph | | | | | No. | Ph | | | |
| DIVISION | Difference of the Exponents Δα | 40 | 0 | Ph | | Af | | 40 | 0 | | | Bf |
| | | 40 | 0 | | | | -Ag | | | | | |
| | actual Division | 41 | 1 | | | | | 41 | 1 | | Be | |
| | | | | | | | | | | | | -Bg |
| | | 43 | 1 ⋮ 18 | Ph, Ae, "R5 | | | | 42 | 2 ⋮17 | u+2=0 | | Bg |
| | | | | | | | | 42 | 2 ⋮17 | u+2=L | | -Bg |
| | | | | | | | | 42 | 2 ⋮17 | | 2Be | |
| | Bf → Bb | 44 | 19 | Ph , Ae | | | | 44 | 19 | | | Bf |
| | Align ← | 45 | 20 | | $Be_0$ = 0 | | -1 | 45 | 20 | $Be_0$ = 0 | 2Be | |
| | | 45 | 20 | Lz, Ae | | | | 45 | 20 | $Be_0$ = L | | Be |

Fig. 18: The microinstructions for division. The multiplier-mantissa in Bf is pushed bit by bit in a shift register (shift left). The multiplicand-mantissa is kept in register Bg.

| Part A (Exponent) | | | | | | | | | | | | | | Notification->B | Part B (Mantissa) | | | | | | | | | | | |
|---|---|---|---|---|---|---|---|---|---|---|---|---|---|---|---|---|---|---|---|---|---|---|---|---|---|---|
| Block 17 (left Part) | | | | | | | | | | Block 15a | | | | | Block 17 (right part) | | | | | | | | Block 15b | | | |
| Operating sequence | | Level | Criterion | | | | | Cond. | to operate | Condition | to operate | Aa | Ab | | Level | Criterion | | | | | Conditio n | to operate | Condition | to operate | Ba | Bb |
| Symb | Description | | No. | Op | S0 | S1 | Ph | | | | | | | | | No. | Op | S0 | S1 | Ph | | | | | | |
| DIVISION (:) | Diffrence of the Exponents Δα | 5 | 40 | : | | | 0 | | Ph | | | Af | | | 5 | 40 | : | | | 0 | | | | | | Bf |
| | | 7 | 40 | : | | | 0 | | | | | | -Ag | | | | | | | | | | | | | |
| | actual Division | 7 | 41 | : | | | 1 | | | | | Ae | | | 9 | 41 | : | | | 1 | | | | | Be | |
| | | | | | | | | | | | | | | | 7 | | | | | | | | | | | -Bg |
| | | 9 | 43 | : | | | 1 ⋮ 18 | | Ph, Ae, "R5 | | | Ae | | | 7 | 42 | : | | | 2 ⋮ 17 | | | u+2=0 | | | Bg |
| | | | | | | | | | | | | Ae | | | 7 | 42 | : | | | 2 ⋮ 17 | | | u+2=L | | | -Bg |
| | | | | | | | | | | | | Ae | | | 6 | 42 | : | | | 2 ⋮ 17 | | | | | 2Be | |
| | Bf → Bb | 7 | 44 | : | | | 19 | | Ph , Ae | | | Ae | | | 5 | 44 | : | | | 19 | | | | | | Bf |
| | Align ← | 6 | 45 | : | | | 20 | | | $Be_0$ = 0 | | Ae | -1 | | 6 | 45 | : | | | 20 | | | $Be_0$ = 0 | | 2Be | |
| | | 2 | 45 | : | | | 20 | | Lz, Ae | | | Ae | | | 6 | 45 | : | | | 20 | | | $Be_0$ = L | | | Be |

Fig. 19: The complete division table with all information columns.

Somewhat puzzling is the fact that the Z3 tested first if the subtraction of Ba and Bb could become negative, during a division, in which case the subtraction was “undone” by using a shortcut bus from Ba to Be (eliminating the result of the subtraction). This extra hardware was not used in the reconstructed Z1 and the non-restoring algorithm seems more elegant than the solution used later in the Z3.

## 10 Input and Output

The input console consists of four columns of ten small plates each. In each column (called Za3, Za2, Za1 and Za0, in that order, from left to right) the operator can pull out any of the digits 0 to 9. He or she can thus enter any decimal number of four digits using the four columns. Pulling a digit plate just generates its binary equivalent in the input console (using four bits). The input console is therefore just a four by ten table of the ten binary equivalents of the digits 0 to 9.

The microcontroller of the Z1 takes then care of passing each decimal digit Za3, Za2, Za1 and Za0 to the datapath through register Ba (at position $Ba_{-15}$ corresponding to the power $2^{-15}$). Za3 is entered first (in register Ba) and then a multiplication by ten takes place. Digit Za2 is next, and another multiplication by ten ensues. This is repeated for all four digits. The binary equivalent of the four decimal digits is contained in the Be after cycle 7. In cycle 8 the mantissa is normalized, if needed. The constant 15 (LLLL in binary) is added once to the exponent (phase 7) in order to account for the fact that the digits are entered at the mantissa bit -15.

The decimal exponent of the number is set using a lever. In cycle 9 as many multiplications with the factor ten are performed, as indicated by the position of the lever for a positive decimal exponent. For a negative decimal exponent $x$ (between -6 and -1) the number is multiplied $(6 + x)$ times with 10 and then with $10^{-6}$ at the end. In this way no multiplications with 0.1 are needed (as was done in the Z3 with additional hardware).

| Part A (Exponent) | | | | | | | | | | | | | | Notification->B | Part B (Mantissa) | | | | | | | | | | | |
|---|---|---|---|---|---|---|---|---|---|---|---|---|---|---|---|---|---|---|---|---|---|---|---|---|---|---|
| Block 17 (left Part) | | | | | | | | | | Block 15a | | | | | Block 17 (right part) | | | | | | | | Block 15b | | | |
| Operating sequence | | Level | Criterion | | | | | Cond. | to operate | Condition | to operate | Aa | Ab | | Level | Criterion | | | | | Condition | to operate | Condition | to operate | Ba | Bb |
| Symb | Description | | No. | Op | S0 | S1 | Ph | | | | | | | | | No. | Op | S0 | S1 | Ph | | | | | | |
| Transform (decimal → dual) (↗) | Ready? 1.<br>Digit x L0L0 2.<br>Digit x L0L0 3.<br>Digit x L0L0 4.<br>Digit | | | | | | | | | | | | | | 12 | 50 | ↗ | | | 0 | u8 | Ph | | | | |
| | | | | | | | | | | | | | | | 4 | 51 | ↗ | | | 1 | | u2, Ph | | | za3 | Be |
| | | | | | | | | | | | | | | | 11 | 52 | ↗ | | | 2 | | Ph | | | 2Be | 8Be |
| | | | | | | | | | | | | | | | 4 | 53 | ↗ | | | 3 | | u2, Ph | | | za2 | Be |
| | | | | | | | | | | | | | | | 11 | 54 | ↗ | | | 4 | | Ph | | | 2Be | 8Be |
| | | | | | | | | | | | | | | | 4 | 55 | ↗ | | | 5 | | u2, Ph | | | za1 | Be |
| | | | | | | | | | | | | | | | 11 | 565 | ↗ | | | 6 | | Ph | | | 2Be | 8Be |
| | | 3 | 57 | ↗ | | | 7 | | | | | | LLLL | | 4 | 57 | ↗ | | | 7 | | u2, Ph | | | za0 | Be |
| | Align ← | 6 | 58 | ↗ | | | 8 | | Ae | | | | | | 4 | 58 | ↗ | | | 8 | | | $Be_0 = 0$ | | 2Be | |
| | | 6 | 58 | ↗ | | | 8 | | | $Be_0 = 0$ | | | -1 | | 4 | 58 | ↗ | | | 8 | | | $Be_0 = L$ | | | Be |
| | | 8 | 58 | ↗ | | | 8 | | | $Be_0 = L$ | Ph | | | | | | | | | | | | | | | |
| | Muliplication with L, 0L | 10 | 59 | ↗ | | | 9 | | Ae | | | | | | 12 | 59 | ↗ | | | 9 | | | | | Be | |
| | | 10 | 59 | ↗ | | | 9 | u6 = L | | | | | LL | | 10 | 59 | ↗ | | | 9 | u6 → | u4 → | | | | ¼ Be |
| | | 10 | 59 | ↗ | | | 9 | u6 = 0 | Ph | | | | | | | | | | | | | | | | | |
| | Align → | 4 | 60 | ↗ | | | 10 | | Ph, Ae | | | | | | 9 | 60 | ↗ | | | 10 | | | $Be_{+1} = 0$ | | Be | |
| | | 8 | 60 | ↗ | | | 10 | | | $Be_{+1} = L$ | | | +1 | | 9 | 60 | ↗ | | | 10 | | | $Be_{+1} = L$ | | | ½ Be |
| | Test $10^{-6}$ | 2 | 61 | ↗ | | | 10 | u7 = 0 | Lz | | | | | | 1 | 61 | ↗ | | | 10 | | u1 ↓ | | | | |
| | | 2 | 61 | ↗ | | | 10 | u7 = L | Ph, Ae<br>Be=>Bg | | | | | | | | | | | | | (delete u8) | | | | |
| | Setting $10^{-6}$ | 1 | 62 | ↗ | | | 11 | | Ph0, S0 (×) | | | | | | | | | | | | | | | | | |

Fig. 20: The microinstructions for decimal-binary conversion. Four decimal digits are entered through a mechanical device.

The table in Fig. 21 shows how to transform a binary number contained in register Bf to a decimal number to be shown in the decimal output panel.

| Part A (Exponent) | | | | | | | | | | | | | | Notification->B | Part B (Mantissa) | | | | | | | | | | | |
|---|---|---|---|---|---|---|---|---|---|---|---|---|---|---|---|---|---|---|---|---|---|---|---|---|---|---|
| Block 17 (left Part) | | | | | | | | | | Block 15a | | | | | Block 17 (right part) | | | | | | | | Block 15b | | | |
| Operating sequence | | Level | Criterion | | | | | Cond. | to operate | Condition | to operate | Aa | Ab | | Level | Criterion | | | | | Condition | to operate | Condition | to operate | Ba | Bb |
| Symb | Description | | No. | Op | S0 | S1 | Ph | | | | | | | | | No. | Op | S0 | S1 | Ph | | | | | | |
| Re-Compile (dual → decimal) (↘) | Setup Operands | 5 | 70 | ↘ | | | 0 | d0 | Ph | | | Af | | | 5 | 70 | ↘ | | | 0 | d0 | | | | | Bf |
| | Setting of $10^{-6}$ | 12 | 71 | ↘ | | | 1 | | Ae | Ae ≤ 3 | Ph | | | B | 12 | 71 | ↘ | | | 1 | | A → | (Ae ≤ 3) | | Be | |
| | | | | | | | | | | Ae > 3 | Ph0, S1, Be (×) => $B_8$ | | | | | | | | | | | | | | | |
| | b - Value 2 Positions → | 12 | 72 | ↘ | | | 2 | | Ph, Ae | | | | | | 10 | 72 | ↘ | | | 2 | | | | | | ¼ Be |
| | | | | | | | | | | | | | | | | | | | | | | | | | | |
| | Muliplication with 10 | 10 | 73 | ↘ | | | 3 | | Ae | Ae < 0 | | | LL | B | 10 | 73 | ↘ | | | 3 | | | A → (Ae < 0) | | | ¼ Be |
| | | | | | | | | | | Ae ≥ 0 | Ph | | | | 12 | (73) | ↘ | | | 3 | | d3 | | | Be | |
| | Align → | 11 | 74 | ↘ | | | 4 | | | Ae ≠ 0 | | | | B | 11 | (74) | ↘ | | | 4 | | | A → (Ae ≠ 0) | | 2Be | |
| | | 9 | 74 | ↘ | | | 4 | | Ae | Ae ≠ 0 | | | -1 | | 9 | | | | | | | | | | | |
| | | | | | | | | | | Ae ≠ 0 | Ph → | | | B | (9) | 74 | ↘ | | | 4 | | | A → (Ae ≠ 0) | | | Be |
| | actual Re-Compile | | | | | | | | | | | | | | 3 | 75 | ↘ | | | 5 | | µPh ,d1 | | | 2Be' | 8Be' |
| | | | | | | | | | | | | | | | 3 | 76 | ↘ | | | 6 | | µPh ,d1 | | | 2Be' | 8Be' |
| | | | | | | | | | | | | | | | 3 | 77 | ↘ | | | 7 | | µPh ,d1 | | | 2Be' | 8Be' |
| | | 3 | 78 | ↘ | | | 8 | | Lz | | | | | | 3 | 78 | ↘ | | | 8 | | d1 ↓<br>(Del → do) | | | | |

Fig. 21: The microinstructions for binary-decimal conversion. Four decimal digits are displayed in a mechanical device.

As said before, the main idea is to take a number such as 0.879 and "expose" its successive digits as the integer before the point, by multiplying the mantissa iteratively with 10. But this has to be done in binary, not in decimal.

The multiplication by 10 is performed in this instruction by multiplying with 8(1+0.25), that is, by raising the exponent by 3 (since $8 = 2^3$) and by taking the mantissa in Be to compute Be+Be/4 (Phase 3).

It is easier to explain the main iteration when the exponent in the register (Ae,Be) is negative. This corresponds to a binary number such as *a*=0.00010 (0.0625 in decimal). But in normalized notation this is $1.0 \times 2^{-4}$. We need to multiply *a* by 10 ($1.01 \times 2^3$ in binary) until the exponent of the number is positive. The number of multiplications needed is two. The binary mantissa 0.00010 transforms after the two multiplications into $0.11001 \times 2^2$, that is, 110.01. The integer part of this binary number is 110, or 6, as it should be for displaying the number. An iterative continuation with 0.01 (0.25 in decimal) will produce a 2 and finally a 5.

The code is somewhat involved since it takes care of avoiding overflow of the mantissa (by moving the whole mantissa at the beginning two bits to the right in Phase 2. The position of the binary point can be displaced two bits to the right in this manner, if we do not change the exponent.

After every integer part of the number is shown in the display (6 in the example above), the integer part of the mantissa Be is erased and we keep the fractional part Be'. The initial four bits are transformed into a decimal digit using a table. Every decimal digit is shown in the output panel (starting with the highest order decimal digit). Every time a multiplication by ten is performed, the exponent arrow in the decimal display displaces one place to the left.

In the case of a positive exponent, at the beginning, the routine multiplies Be by $10^{-6}$ using a "call" to the multiplication routine. In Phase 1, if Ae>3, the opcode is changed to the opcode for multiplication, and control transfers to the special portion of the multiplication instruction that multiplies with $10^{-6}$. Control returns back to this instruction afterwards. In this way the exponent Ae is made negative, and we are back to the previous case. The Z1 has to keep a record of this, and this is done by advancing or moving backwards the decimal point display lever in front of the display unit. For example, using this approach, the decimal number 687 is transformed into 0.000687. The decimal point stands at $-6$. After six multiplications with 10, we obtain the successive digits 6,8,7, and every multiplication advances the decimal point +1 places. At the end, the display shows 687 with decimal exponent zero. It is a curious way of handling numbers with positive exponents, but this allows the Z1 to avoid division by 10, which would be the alternative for getting the digits of 687, one after the other. The Z3 used division by 10 and did not need the special call to multiplication by $10^{-6}$.

## 11 The mechanical implementation

As explained before, the datapaths for exponent and mantissa are implemented in 12 mechanical layers. A microinstruction activates one layer, the left or right part. For the exponent, the left side defines the loading of register Aa, and the right side of register Ab. Fig. 22 shows the mechanical layers of the exponent part and their relationship to the microinstruction plates. Fig. 23 is the equivalent picture for the mantissa part.

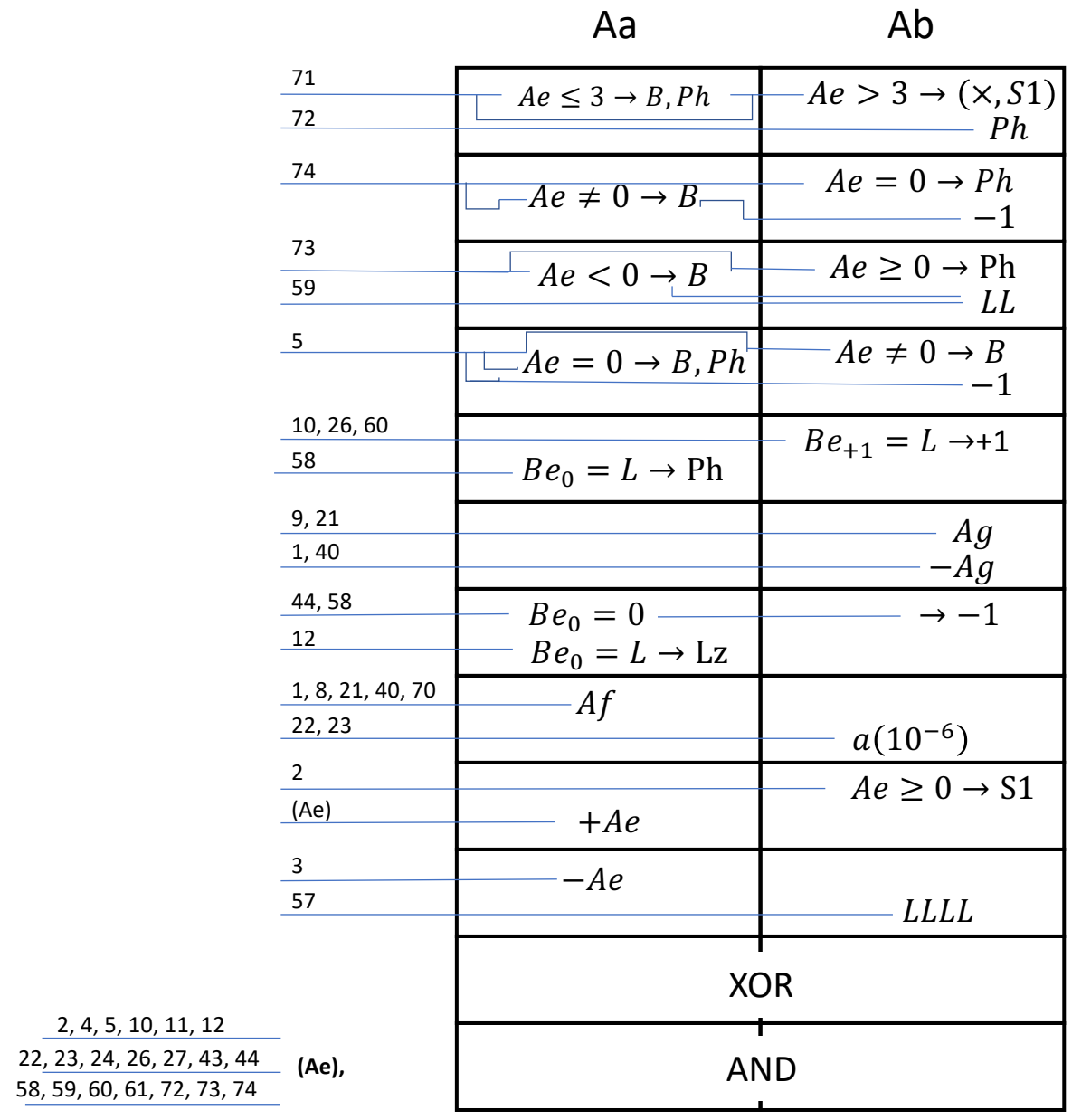

**12** Loop for binary-decimal conversion

**11** Normalization loop for binary-decimal conversion

**10** Loop for multiplication by ten

**9** Loop for mantissa shift of smaller number in addition

**8** Loop for mantissa normalization after addition

**7** Selection of positive or negative Ag

**6** Loop for mantissa normalization

**5** $Af \rightarrow Aa$; exponent for multiplication by $10^{-6}$

**4** Circulation of Ae; S1 =1 if mantissa difference < 0

**3** -Ae->Aa, and 15 to Ab

**2** Parallel bitwise XOR of the bits of Aa and Ab

**1** Parallel bitwise AND of the bits of Aa and Ab

Fig. 22: The microcode activated actions in the exponent processor stack. The figure shows the criterion numbers (on the left) that activatesone or the other side, in a stack of 12 mechanical layers. The layers 1 and 2 are reserved for the addition unit (XOR, and AND bitwise computation).

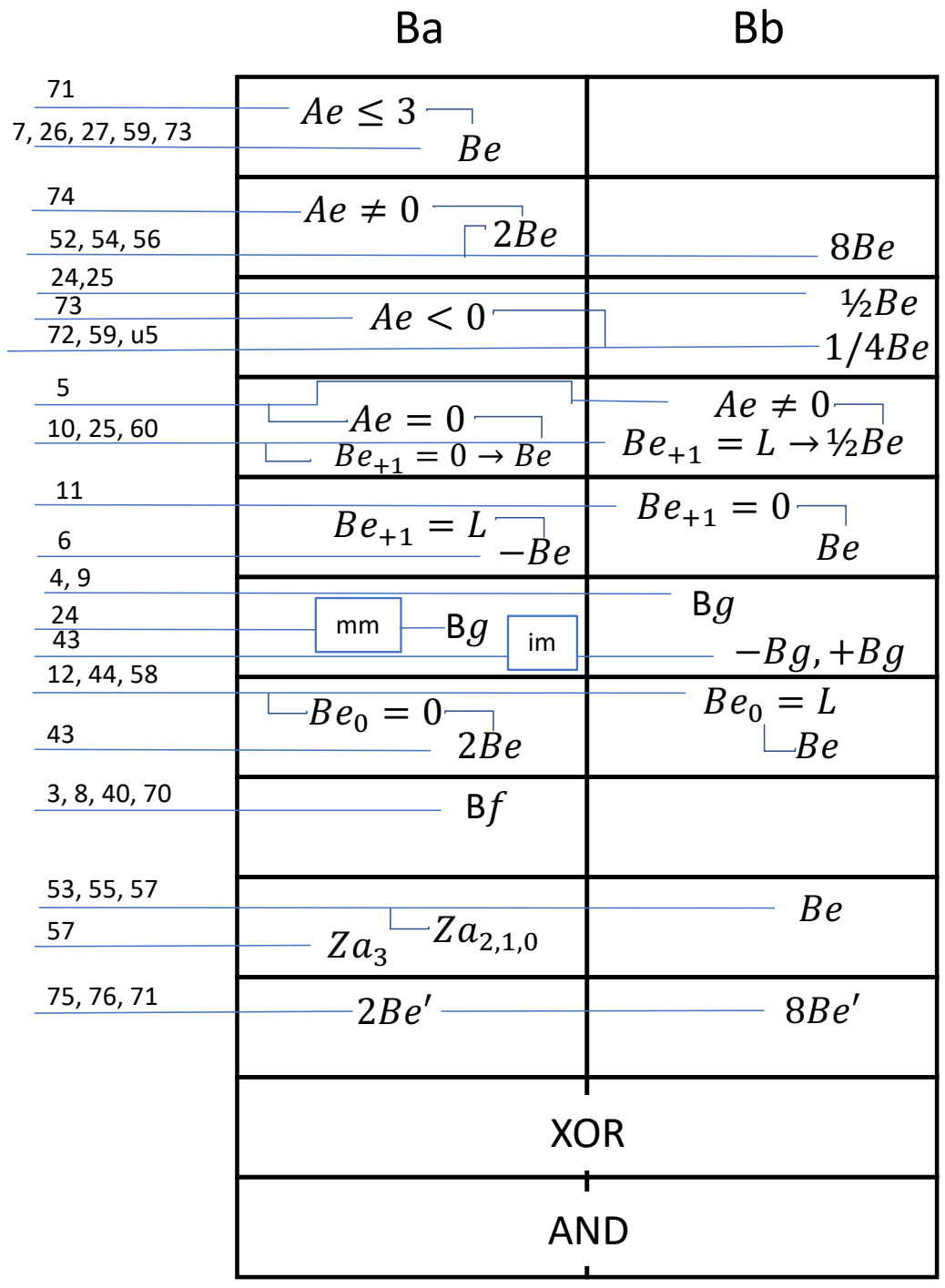

**12** $Be \rightarrow Ba$; Loop binary-decimal, mult. with $10^{-6}$ .

**11** Normalization; multiplication by ten loop

**10** Part of multiplication; Multiplication by ten loop

**9** Loop for mantissa shift of smaller number in addition

**8** Loop for mantissa normalization after addition

**7** Selection of $+Bg$ in mult.; of $\pm Bg$ in division

**6** Loop for mantissa normalization

**5** $Bf \rightarrow Ba$

**4** Entry of digits for decimal-binary conversion

**3** Multiplication by 10 of the rest mantissa

**2** Parallel bitwise XOR of the bits of Ba and Bb

**1** Parallel bitwise AND of the bits of Ba and Bb

Fig. 23: The microcode activated actions in the mantissa processor stack. The figure shows the criterion number which activates one or the other side, in a stack of 12 mechanical layers. The layers 1 and 2 are reserved for the addition unit (XOR, and AND bitwise computation).

For example, layer 5 of the mantissa part (Fig. 23) is activated by the microinstruction plates 3, 8, 40, 70. They all need register Bf to be loaded to register Ba. Sometimes several plates are needed to start the same operation, because the instruction opcodes can be different. For example, microinstruction plate 3 is used for addition, while microinstruction plate 40 is used for multiplication. These microinstructions they all need Bf in Ba.

In Figs. 22 and 23, a signal from the exponent layers to "B", means that a signal is sent from Fig. 22 to Fig. 23. This is summarized in Fig.24, where we can see, for example, how the condition Ae<=3 closes a relay for each bit of the mantissa, sending Be to Ba. The condition Ae=/0 send 2Be to Ba. A mechanical connection between the exponent part and the mantissa part of the processor is activated. Figs. 22, 23, and 24, together, provide a very clear picture of the 12 layers of the Z1 and their interconnections.

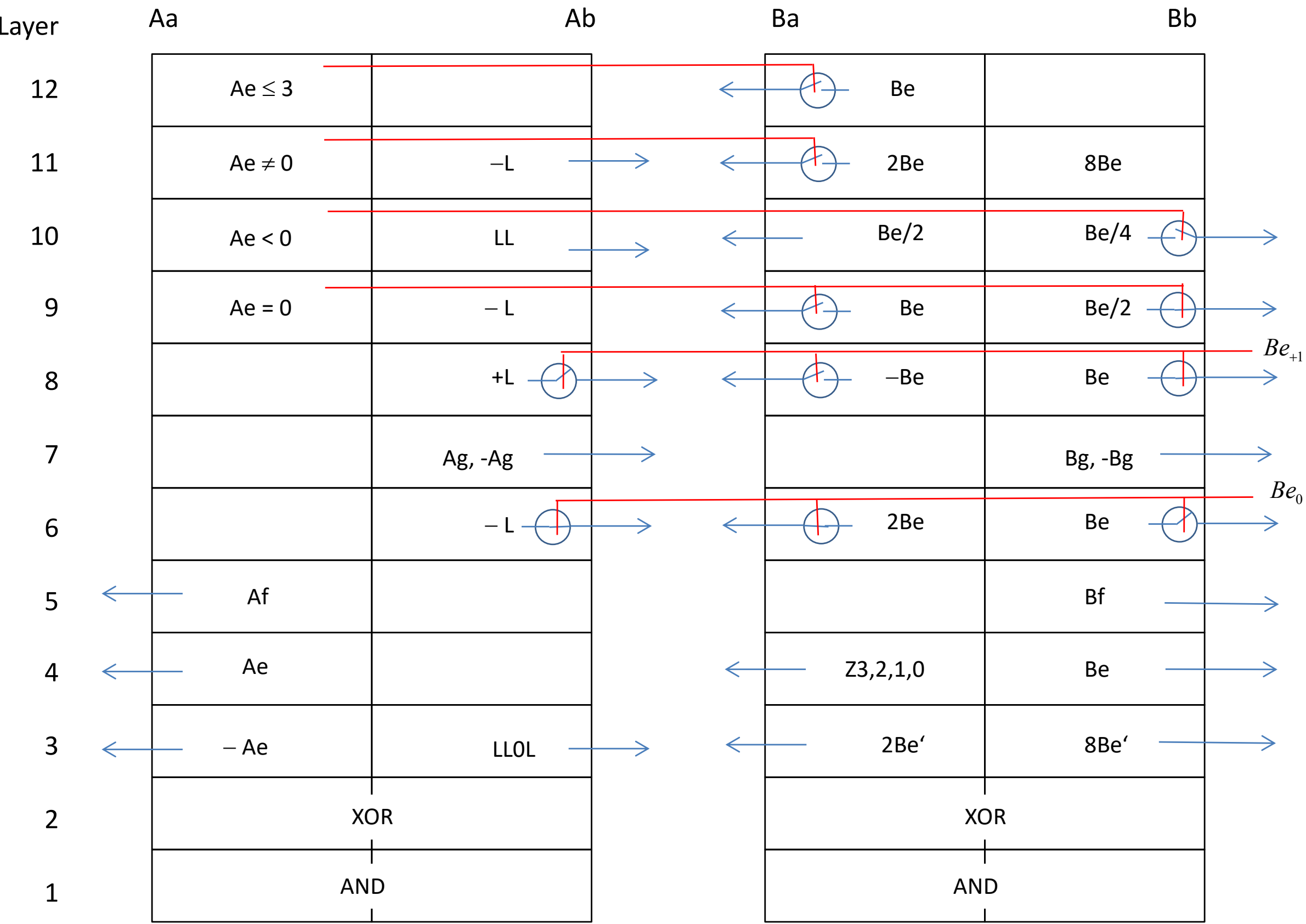

Fig. 24: Communication between the exponent and the mantissa ALUs.

## 11 Conclusions

The original Z1 was destroyed during an allied bombing raid flown in December 1943 over Berlin. It is impossible to decide today if the original Z1 was identical to the reconstructed Z1. The few photographs that survived show that the original was bulkier and had a less "regular" form. Here we can only take Zuse at his word. However, I think that he had no real reason to consciously "embellish" the original machine through the reconstruction. Memory can be a tricky fellow though. The few notes Zuse scribbled between 1935 and 1938 seem to be consistent with the later reconstruction. The Z3 was finished in 1941, and according to Zuse, it was very similar in design.

Siemens (the company that acquired Zuse's computer company) financed the reconstruction of the Z1 in the 1980s. Zuse did all the construction work at his home, having two students assisting him. When the Z1 was finished, part of the wall in the upper floor of Zuse's home had to be removed, so that a large crane could lift the machine for its transportation to Berlin.

The reconstructed Z1 is a very elegant computer, consisting of thousands of components but not one too many. It would have been possible to use only two shifters at the output of the mantissa ALU (a shift by one bit to the left, and one bit to the right), but the selection of shifters made by Zuse speeds up the basic arithmetical operations significantly, at a low cost in components. I find the processor of the Z1 rather more elegant than the processor of the Z3, since it is more compact and "fundamental". It is as if when Zuse moved to telephone relays, the simpler and more reliable components allowed him to be "profligate" with the size of the CPU. The same happened years after the Z3, when the Z4 was finished. The Z4 was just a larger Z3 with a larger instruction set, but the computer architecture was roughly the same, even though the Z4 had more instructions. The mechanical Z1 never worked consistently and Zuse himself later called the mechanical realization "a dead end". He used to joke that the 1989 reconstruction of the Z1 was quite accurate, because the original was not reliable, and neither was the reconstruction. Curiously, the mechanical memory design was sufficiently dependable to use it again for the Z4, as a way of saving telephone relays. The mechanical memory of the Z4 was operational from 1950 to 1955 in Switzerland, where the machine was installed at the ETH Zürich [7].

What I find most surprising is how the young Konrad Zuse could come to such an elegant design for a computing engine. Whereas the ENIAC, or Mark I teams in the US consisted of seasoned scientists and electronic experts, Zuse was working in isolation and without real previous experience. From the architectural point of view, we compute today as Zuse did in 1938, not as the ENIAC did in 1945 [8]. More elegant architectures were only introduced later with the EDVAC report and the bit-serial machines developed by von Neumann and Turing. John von Neumann lived from 1926 to 1929 in Berlin and was the youngest Lecturer (Privatdozent) at the University of Berlin. Konrad Zuse and von Neumann may have crossed paths inadvertently during those years. So much could have happened in Berlin before madness took over and a dark night fell over Germany.

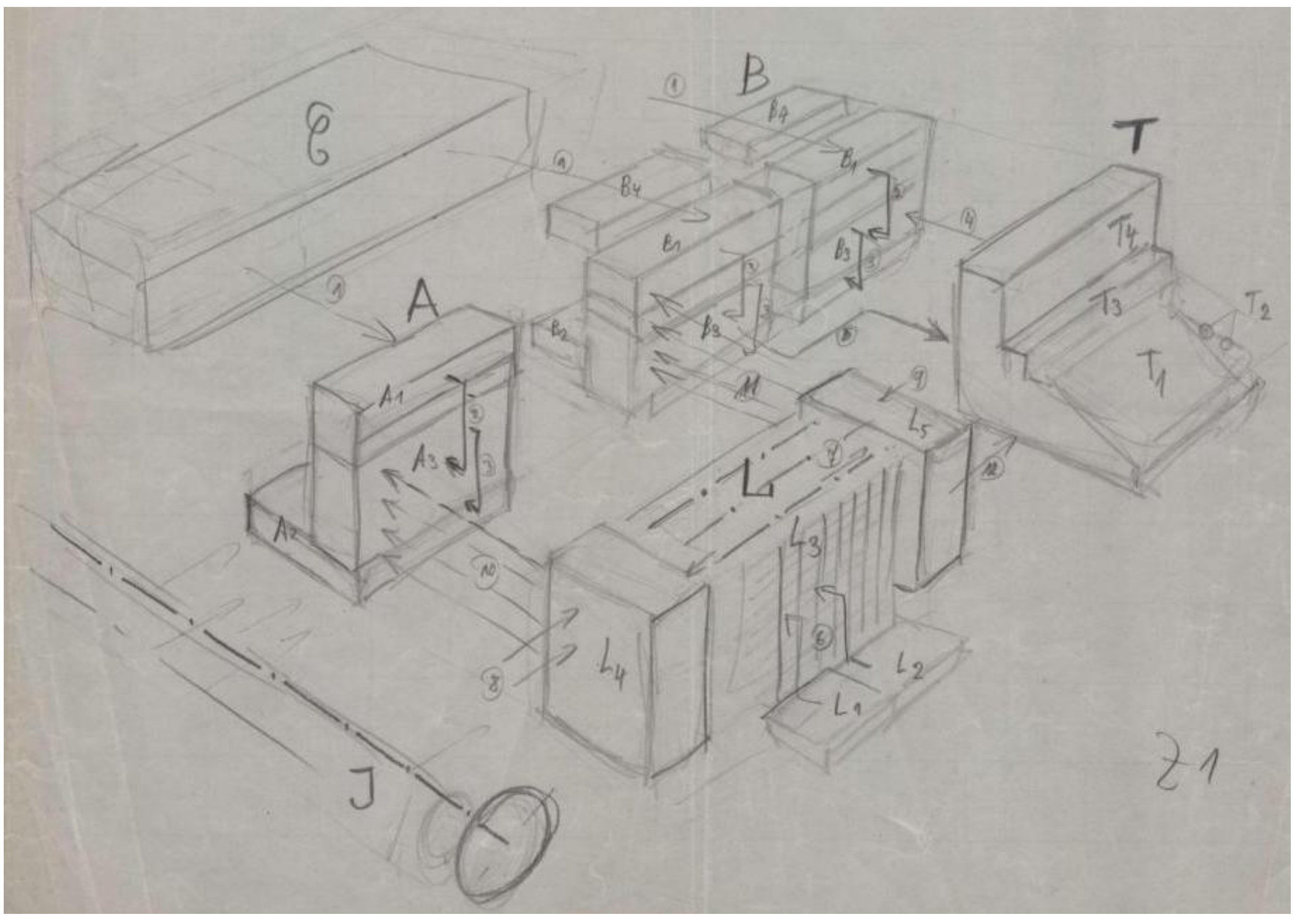

Fig. 20: Sketch of one of Zuse's early designs for a Z1 reconstruction. Undated.